\documentclass[man,floatsintext]{apa7}
\usepackage{IEEEtrantools}
\usepackage[utf8x]{inputenc}
\usepackage{setspace}
\usepackage{amsmath,amsthm,amssymb}
\usepackage{graphicx}
\usepackage{bm}
 \usepackage[backend=biber, style=apa, citestyle=apa]{biblatex}
\usepackage[colorinlistoftodos]{todonotes}
\usepackage[flushleft]{threeparttable}
\usepackage{booktabs,caption}
\usepackage{colortbl}
\usepackage{graphicx}
\usepackage{pdflscape}
\usepackage{caption}
\usepackage{subcaption}
\usepackage{enumitem}
\usepackage{graphicx}
\usepackage{bm}
\usepackage{textcomp}
\usepackage{textgreek}
\usepackage{caption}
\usepackage{multicol}
\usepackage{IEEEtrantools}
\usepackage{booktabs}
\usepackage{authblk}
\usepackage{outline}
\usepackage{mathrsfs}
\usepackage{euscript}
\usepackage{amsbsy}
\usepackage{mathtools}
\usepackage{multirow}
\usepackage{rotating}
\usepackage[ruled,vlined]{algorithm2e}
\SetKwRepeat{Do}{do}{while}%
\usepackage{array}
\usepackage{comment}
\usepackage{indentfirst}

\newcommand{\vect}[1]{#1}
\newcommand{\Z}{\mathbb{Z}}
\newcommand{\R}{\mathbb{R}}
\newcommand{\vecop}{\operatorname{vec}} 
 
\newcommand{\diag}{\operatorname{diag}}
\def\CC{\mathbb C}
\newcommand{\Nzero}{\mathbb{N}_0}

\DeclareMathOperator{\Cov}{\mathbb{C}ov}

\DeclareMathOperator{\E}{\mathbb{E}}

\DeclareMathAlphabet{\mathscrbf}{OMS}{mdugm}{b}{n}
\newtheorem*{@remark}{\bf Remark}

\newcommand{\Prob}{\operatorname{P}}

\theoremstyle{definition}

\newcommand{\acmfXhat}{\widehat{\bm{\Gamma}}_{X}}
\newcommand{\acmfX}{\bm{\Gamma}_{X}}
\newcommand{\acmfZhat}{\widehat{\bm{\Gamma}}_{Z}}
\newcommand{\acmfZ}{\bm{\Gamma}_{Z}}

\newcommand{\Dim}{d} 

\newcommand{\CI}[2]{[\,#1,\ #2\,]}

\usepackage{mathtools}
\mathtoolsset{showonlyrefs}

\title{Analytic Standard Errors for Latent Gaussian Discrete-Valued Multivariate Time Series}

\shorttitle{}

\authorsnames[1,2,3,1,1,1]{Christopher M. Crawford, Marie-Christine D\"uker, Younghoon Kim, Vladas Pipiras, Barbara L. Fredrickson, Zachary F. Fisher}
\authorsaffiliations{{University of North Carolina at Chapel Hill}, {Technical University of Munich}, {Cornell University}}

\authornote{
\setlength{\parindent}{0pt}

\textbf{Corresponding author:} Christopher M. Crawford, University of North Carolina at Chapel Hill, Chapel Hill, North Carolina, United States. Email: cmcrawford4792@gmail.com

\noindent \textbf{Funding statement:} This research received no specific grant funding from any funding agency, commercial or not-for-profit sectors.

\noindent \textbf{Competing interests:} The authors declare no competing interests.
}

\abstract{Unlike their continuous-valued counterparts, there are no universally preferred methodologies for modeling discrete-valued time series. This is especially problematic in fields such as psychology and education, where repeated-measures data often take the form of count, dichotomous, and ordered categorical variables. To address the need for flexible methodology for analyzing discrete-valued time series data, a copula-style multivariate model defined through deterministic functions of a latent stationary Gaussian vector series has been proposed. This model has several promising features, including the ability to accommodate a wide variety of marginal distributions within the same model while also allowing for the most flexible autocorrelation structure possible. We extend this framework by deriving analytic standard errors to facilitate inference on the latent Gaussian dynamics. In so doing, we establish the joint asymptotic normality of estimators of the parameters governing the latent Gaussian series and the marginal distributions. The performance of these analytic standard errors is examined in a simulation study and an empirical application.}


\begin{document}
\makeatletter
\@ifundefined{@affil}{\def\@affil{~}{}}
\makeatother
\maketitle

Across the social, behavioral, and health sciences, interest in the study of dynamic processes has grown rapidly. This increased focus, coupled with the near ubiquity of mobile and wearable devices, has resulted in the widespread collection of multivariate time series data. One common approach for analyzing such data is the vector autoregressive (VAR) model, which describes within-person dynamics as lagged associations between constituent variables \parencite{Lutkepohl2005}. Recent work, for example, has employed VAR-based methods to investigate a range of processes, including psychopathology \parencite{Bringmann2013}, emotion dynamics \parencite{Pe2015}, and personality development \parencite{Beck2020}.

A key assumption of the VAR model is that all variables are measured on a continuous scale, such that the outcomes of the process are real numbers. In many applications, however, researchers are interested in processes characterized by discrete-valued variables, such as nonnegative integer counts or binary indicators. For example, recent work has examined processes consisting of counts of drinks \parencite{Soyster2022, Liu2025, DeMartini2022}, total daily social interactions \parencite{Elmer2025}, the presence or absence of distinct mood states \parencite{Fisher2020, Howe2020}, ratings of parent–child interaction quality \parencite{VanKeer2019}, and digital phenotyping indicators \parencite{Jacobucci2024, Ram2020}.

Despite the prevalence of discrete-valued processes, modeling these dynamics presents a number of statistical and computational challenges. For example, discrete-valued processes convey a limited amount of information per observation compared to their continuous-valued counterparts. As such, it is more difficult to detect temporal dependence and estimate model parameters precisely. This is further complicated by processes that exhibit overdispersion or zero  inflation---features not well represented by common marginal distributions. Consequently, there is no single predominant approach for modeling multivariate discrete-valued series, in contrast to the well-established VAR framework for continuous-valued processes.

A number of approaches have been developed to address the challenges inherent in modeling discrete-valued series. In the univariate setting, early approaches were based on the autoregressive moving average (ARMA) framework. The integer autoregressive (INAR) model, for example, is constructed via a thinned autoregressive (AR) process \parencite{Jacobs1978}. However, this approach, along with the discrete AR (DAR) model, is unable to produce negative autocorrelations, thereby limiting its utility. More recent work has addressed this shortcoming within the broad frameworks of generalized state-space models and Bayesian dynamic generalized linear models \parencite[for a review, see][]{Davis2021}. These approaches are not without limitations, however. In the generalized state-space framework, for example, the observed discrete-valued process is specified as conditional on the unobserved state; as a result, the marginal distribution of the series can deviate substantially from the assumed observation family \parencite{Benjamin2003}.

Methods for modeling multivariate discrete-valued processes have received comparatively less attention \parencite[for a review, see][]{Fokianos2024}. Notable approaches largely belong to the class of parameter-driven models \parencite{Cox1981}, which includes generalized dynamic factor models \parencite[e.g.,][]{Cui2014, Zhang2007, Jung2011} and generalized state-space models \parencite[e.g.,][]{Jorgensen1999, Vanrijn2008}. Importantly, these frameworks typically posit an exponential-family observation model for the discrete-valued process, which does not accommodate zero inflation. Recent work by Henry and colleagues (\citeyear{Henry2024}) relaxes this requirement by permitting flexible observation models within a state-space framework; at present, however, their implementation is specific to ordinal outcomes. Related work by Asparouhov et al. (\citeyear{Asparouhov2018}) models discrete-valued series within the dynamic structural equation modeling (DSEM) framework \parencite[see also][]{McNeish2023}. However, the probit-link construction proposed in this approach does not extend to many common discrete outcomes, such as counts. 

An alternative class of models employs a latent Gaussian process and a copula-style distributional transformation. That is, the observed discrete-valued series is modeled as a function of a latent Gaussian process, which follows a parametric form of interest. Notably, this approach overcomes each of the noted challenges by allowing for both a general correlation structure and the specification of any marginal distribution. In the univariate setting, Jia and colleagues (\citeyear{Jia2023}) proposed a number of procedures for estimating the latent Gaussian AR dynamics, including Gaussian pseudo-likelihood and implied Yule--Walker estimators, in addition to likelihood estimation via particle filtering and sequential Monte Carlo methods. Recent work has extended this approach to contexts in which the latent Gaussian series follows a dynamic factor model \parencite{Kim2025} and a high-dimensional VAR process \parencite{Duker2024}. Though estimation of the latent Gaussian dynamics in these settings has been established, the construction of valid standard errors for the effects of interest has not yet been addressed. Indeed, the asymptotic properties of the noted estimators, such as normality, have not been explored \parencite{Jia2023}.   

The current article makes three contributions to the existing literature on this class of models. First, we derive closed-form expressions for the asymptotic standard errors of the estimated latent Gaussian dynamics under Yule--Walker estimation, facilitating uncertainty quantification for the effects of interest. Second, we establish the asymptotic normality of the Yule--Walker estimator. Third, we implement the Yule--Walker estimator and derived standard errors in the open-source R package \texttt{timecop} \parencite{timecop}. We restrict our focus in the current article to the Yule--Walker estimator under the supposition that the latent Gaussian series follows a causal VAR process. Though likelihood estimation is assumed to be asymptotically normal \parencite{Jia2023}, this approach is computationally intensive due to the presence of high-dimensional integrals, and methods for approximating these integrals are unwieldy for long time series lengths \parencite[cf.][]{Nguyen2025}. As such, likelihood-based inference may be infeasible in many applied scenarios. 

The remainder of the article is as follows. First, we describe the specification of the latent Gaussian process via a copula-style distributional transformation, and we establish the relation between the covariance structures of the observed and latent series. Next, we detail the Yule--Walker estimation procedure for the latent Gaussian dynamics. We then show that the Yule--Walker estimator is asymptotically normal, and we derive closed-form expressions for the asymptotic standard errors of the estimated latent Gaussian dynamics. Finally, the performance of these standard errors is assessed in both a simulation study and empirical example. 

\section{Model Specification and Estimation}

The current approach considers a strictly stationary multivariate discrete-valued time series model. Let $\{\vect{X}_t\}_{t \in \Z}$ be an observed $d$-dimensional discrete-valued process, such that $X_{i,t} \in \Nzero := \{0,1,2, \dots \}$. To ensure that the $i^{th}$ component series, $\{X_{i,t}\}$, has a specified marginal distribution $F_i(x) = \Prob[X_{i,t} \leq x]$ for each $t$, we employ a copula-style transformation: 
\begin{equation}
    X_{i,t} = G_i(Z_{i,t}), 
    \hspace{0.2cm}
    G_{i}(z_{i}) = F_{i}^{-1} \bigl(\Phi(z_{i}) \bigr),
    \hspace{0.2cm}
    G(z) = \bigl(G_{1}(z_{1}),\dots, G_{d}(z_{d}) \bigr)', 
    \hspace{0.2cm}
    z \in \R^{d},
    \label{eq:marginal}
\end{equation}
where $\Phi(\cdot)$ denotes the standard Gaussian cumulative distribution function (CDF), $F_{i}^{-1}(u) = \inf\{x ~|~ F_{i}(x) \geq u\}, u \in (0,1)$, is the generalized inverse (quantile function) of the $i^{th}$ marginal distribution function. The marginal distribution $F_i$ is parameterized by $\theta_i$, which may be a scalar or vector, and $\{\vect{Z}_t\}_{t \in \Z}$ is a stationary latent $d$-dimensional standard Gaussian process with $\E[Z_{i,t}] = 0$ and $\E[Z_{i,t}^2] = 1$. We further denote the lag-$h$ autocovariance (ACVF) and autocorrelation (ACF) functions of $\{\vect{Z}_t\}$ as $\Gamma_{Z}(h)$ and $R_{Z}(h)$, respectively, such that $\Gamma_{Z}(h) = R_{Z}(h) = \E [\vect{Z}_{t+h} \vect{Z}_{t}']$. Thus, the discrete-valued series, $\{\vect{X}_t\}$, is modeled as a deterministic function of the latent standard Gaussian process, $\{\vect{Z}_t\}$. 

This approach to constructing $\{\vect{X}_t\}$ has a number of benefits. First, the copula-style distributional transformation detailed in \eqref{eq:marginal} can accommodate any marginal distribution, and this construction ensures that the marginal CDF of $X_{i,t}$ is $F_i$. Second, the components of the resulting process need not have the same marginal distribution. That is, the observed series can consist of mixed variables. Finally, temporal dependence in $\{\vect{Z}_t\}$ is inherited by $\{\vect{X}_t\}$, allowing for estimation and inference to be conducted via procedures that rely only on the second-order properties of the latent Gaussian series.  

\subsection{Relationship Between Covariance Structures}

The copula-style distributional transformation in \eqref{eq:marginal} describes a deterministic relation between $\{\vect{X}_t\}$ and $\{\vect{Z}_t\}$. In the continuous case, the joint distribution of the observed process can be derived via the typical change-of-variables approach, allowing for likelihood-based estimation \parencite[e.g.,][]{Andersson2025}. When $\{\vect{X}_t\}$ is discrete, however, this procedure is not applicable. Instead, we rely on a Hermite expansion of the distributional transformation in \eqref{eq:marginal}, from which a similar relation between the covariance structures of the observed and latent processes can be derived \parencite[see][Ch.~5]{Pipiras2017}. We define the lag-$h$ ACVF of $\{\vect{X}_t\}$ as $\Gamma_{X}(h) = \E[\vect{X}_{t+h} \vect{X}_{t}'] - \E[\vect{X}_{t+h}] \E[\vect{X}_{t}]'$, with individual entries of this matrix function denoted as $\Gamma_{X,ij}(h)$ for $i,j = 1, \dots, d$. A Hermite expansion for the component functions of $G$ in \eqref{eq:marginal} is defined as 
\begin{equation}
    \label{eq:Hermiteexpansion}
    G_{i}(z) = \sum_{k =0}^{\infty} \frac{c_{i,k}}{k!} H_{k}(z),
\end{equation}
where $H_{k}(z)$ is the $k$th-order Hermite polynomial:
\begin{equation}
    \label{eq:polynomial}
    H_{k}(z) = (-1)^{k} e^{z^2/2} \frac{\partial^{k}}{\partial z^{k}} e^{-z^2/2},
\end{equation}
and $c_{i,k}$ are the Hermite coefficients:
\begin{equation} 
\label{eq:Hermitecoeff}
c_{i,k} = \E \bigl[G_{i}(Z_{i,0})H_{k}(Z_{i,0}) \bigr].
\end{equation}
Under mild regularity conditions, the Hermite coefficients in \eqref{eq:Hermitecoeff} satisfy the following representation \parencite{Jia2023}:
\begin{equation}
    \label{eq:representation_coef}
    c_{i,k} = \frac{1}{\sqrt{2\pi}} 
    \sum_{n =0}^{\infty} e^{ -Q_{i,n}^2/2} H_{k-1}( Q_{i,n} ), \hspace{0.2cm}
    k \geq 1, 
\end{equation}
with $Q_{i,n} = \Phi^{-1}(C_{i,n})$ and $C_{i,n} = \Prob[X_{i,t} \leq n]$. Thus, $c_{i,k}$ depends on the marginal distribution of the $i$th component of $\{\vect{X}_t\}$. When $Q_{i,n} = \pm \infty$ (i.e., $C_{i,n} = 0$ or 1), the summand in \eqref{eq:representation_coef} is interpreted as 0. For many marginal distributions of interest, this occurs at moderate values of $n$. In practice, therefore, this summation can be computed using a relatively small number of terms. 

The relationship between $\Gamma_{X}(h)$ and $\Gamma_{Z}(h)$ \parencite[see][Ch.~5]{Pipiras2017} can then be written as 
\begin{equation} 
\label{eq:cov}
\Gamma_{X}(h)  = \left( \sum_{k=1}^{\infty} 
\frac{c_{i,k} c_{j,k}}{k!} \Gamma_{Z,ij}(h)^k \right)_{i,j = 1, \dots, d}, 
\end{equation}
where $c_{i,k}$ and $c_{j,k}$ are the $k$th-order Hermite coefficients corresponding to entry $(i,j)$ of $\Gamma_{X}(h)$. Prior work has demonstrated that truncation of the sum in \eqref{eq:cov} at moderate values of $k$ is sufficient in practice \parencite{Jia2023}. We can concisely represent this relation using the following link function:
\begin{equation} 
\label{eq:function_ell}
\ell(u) = \bigl( \ell_{ij}(u) \bigr)_{i,j = 1, \dots, d}
\hspace{0.2cm} \text{ with } \hspace{0.2cm} 
\ell_{ij}(u) = \sum_{k=1}^{\infty} \frac{c_{i,k} c_{j,k}}{k!} u^k.
\end{equation}
This construction admits a number of notable features. First, $\ell_{ij}$ is strictly increasing \parencite[see][]{Duker2024}, and it maps $[-1,1]$ into $[\ell_{ij}(-1), \ell_{ij}(1)]$. Most notably, since $\ell_{ij}$ is strictly increasing, so is its inverse, $\ell^{-1}_{ij}$, a point to which we will return shortly. Next, the representation described herein produces the most flexible covariance structure possible for two random variables with the same marginal distribution \parencite[see][]{Jia2023}. Negative autocovariances are, therefore, easily accommodated, in contrast to the previously noted DAR and INAR models. Finally, the link function detailed in \eqref{eq:function_ell} depends only on the parameters of the marginal distributions comprising the Hermite coefficients in \eqref{eq:Hermitecoeff}. Thus, estimation of the specified marginal parameters allows for estimation of the link functions and their inverses.

\subsection{Estimation of the Inverse Link Function}

The link function in \eqref{eq:function_ell} specifies a deterministic relation between the observed and latent autocovariances, and results from Jia et al. (\citeyear{Jia2023}) in \eqref{eq:representation_coef} provide an approach for feasible computation of the Hermite coefficients. Two challenges, however, remain. First, the link function depends on the marginal distributions of the discrete-valued process, $\{\vect{X}_t\}$. That is, $\ell_{ij}$ is a function of $F_i$ and $F_j$, the marginal distributions of $\{X_{i,t}\}$ and $\{X_{j,t}\}$, respectively, which are parameterized by  $\theta_i$ and  $\theta_j$. For example, if $\{X_{i,t}\}$ is a Bernoulli process, then $\theta_i = p_i$, the probability of observing a 1. Thus, an estimate of $\ell_{ij}$ is needed, as the parameters of the constituent marginal distributions are rarely known in practice. As $\ell_{ij}$ depends only on these marginal distributions, estimation of the link function follows directly from estimation of the corresponding marginal parameters, which can proceed through method-of-moments estimation. In the Bernoulli case, for example, $p_i$ is identified by the moment condition $\E[X_{i,t} - p_i] = 0$, such that the estimator is the sample mean of the observed data. These estimates can then be used to compute the Hermite coefficients in \eqref{eq:representation_coef} and construct the estimated link function. 

Second, the relation detailed in \eqref{eq:cov} is specified in terms of the autocovariances of the latent Gaussian process, $\{\vect{Z}_t\}$; that is, $\Gamma_{X,ij}(h)$ is a function of $\Gamma_{Z,ij}(h)$. However, the second-order properties of the latent Gaussian process are needed to proceed with Yule--Walker estimation of the dynamics of interest. To obtain these autocovariances, we therefore define the following inverse link function:
\begin{equation}\label{eq:inverse}
  \Gamma_{Z,ij}(h) = g_{ij} \bigl(\Gamma_{X,ij}(h) \bigr),
  \qquad g_{ij}(v) := \ell_{ij}^{-1}(v),
\end{equation}
where the construction of $g_{ij}$ is justified via the strictly increasing nature of $\ell_{ij}$. To compute $g_{ij}$, we consider a partition of the interval $[-1,1]$, the domain of $\ell_{ij}$, by $u_0, u_1, \dots, u_M$. The points $(v_m, u_m) = (v_m, g_{ij}(v_m))$ for the relation $g_{ij}(v) = u$ are then obtained by setting $v_m = \ell_{ij}(u_m)$. Points not included in the partition of $[-1,1]$ are obtained via natural cubic spline interpolation. More specifically, we define $\tilde{g}_{ij}(v)$ on $[v_0,v_M]$
as a piecewise cubic function:
\begin{equation}
\tilde{g}_{ij}(v)
=
\sum_{m=1}^{M}\tilde{g}_{ij,m}(v)\,\mathbf{1}_{[v_{m-1},v_m)}(v),
\nonumber
\end{equation}
where, for $v \in [v_{m-1},v_{m})$, $h_m:=v_m-v_{m-1}$, and $m = 1, \dots M$,
\begin{equation}
\tilde{g}_{ij,m}(v)
=
a_{m-1}\frac{(v_m-v)^3}{6h_m}
+a_m\frac{(v-v_{m-1})^3}{6h_m}
+b_{1,m}(v-v_{m-1})
+b_{2,m}(v_m-v).
\nonumber
\end{equation}
The coefficients $\{a_m\}_{m=0}^{M}$ and $\{b_{1,m},b_{2,m}\}_{m=1}^{M}$ are chosen so that
$\tilde{g}_{ij}$ is the natural cubic spline interpolant of the knots $(v_m,u_m)$; that is,
it satisfies the standard interpolation and smoothness conditions, such as being twice continuously differentiable at the knots. These requirements yield a symmetric tridiagonal linear system, which can be solved in the usual manner \parencite[e.g.,][]{Kress1998}. Finally, we extend $\tilde{g}_{ij}$ to all $v\in[-1,1]$ by
\begin{equation}
\widehat{g}_{ij}(v)=
\begin{cases}
-1, & v\le v_0,\\
\tilde{g}_{ij}(v), & v_0 < v < v_M,\\
1, & v\ge v_M,
\end{cases}
\nonumber
\end{equation}
where $\widehat{g}_{ij}$ is the resultant estimated inverse link function. For more information regarding this procedure, as well as graphical representations of the interpolated inverse link function for several marginal distributions, see Kim et al. (\citeyear{Kim2025}).

\subsection{Estimation of Latent Gaussian Dynamics}

In the current article, we consider the case where the latent Gaussian series, $\{\vect{Z}_t\}$, follows a VAR process of order $p$, VAR($p$), such that
\begin{equation}
    \label{eq:VAR}
    \vect{Z}_t = \sum_{u = 1}^p \Phi_u \vect{Z}_{t-u} + 
    \epsilon_t,
\end{equation}
where $\Phi_1, \dots, \Phi_p$ are $d \times d$ transition matrices, and $\epsilon_t \sim WN(0, \Sigma_{\epsilon})$ is a Gaussian white noise process with $\E[\epsilon_t] = 0$ and $\E[\epsilon_t \epsilon'_{s}] = 0$ for $s \neq t$. We further assume that the VAR($p$) process is causal, such that $\det(\Phi(z)) \neq 0$ for $|z| \leq 1, z \in \CC$, where $\Phi(z) = I_{d} - \Phi_{1}z - \dots - \Phi_{p}z^{p}$. The VAR($p$) process in \eqref{eq:VAR} can be written in the familiar linear regression form as 
\begin{equation} 
    \label{eq:VARlinear}
    \underbrace{\begin{pmatrix}
        \vect{Z}_{p+1}' \\
        \vdots \\
        \vect{Z}_{T}'
    \end{pmatrix}}_{\mathcal{Y}_{Z}}
    =
    \underbrace{\begin{pmatrix}
        \vect{Z}_{p}' & \dots & \vect{Z}_{1}' \\
        \vdots & \ddots & \vdots \\
        \vect{Z}_{T-1}' & \dots & \vect{Z}_{T-p}'
    \end{pmatrix}}_{\mathcal{X}_{Z}}
    \underbrace{\begin{pmatrix}
        \Phi_{1}' \\
        \vdots \\
        \Phi_{p}'
    \end{pmatrix}}_{B}
    +
    \underbrace{\begin{pmatrix}
        \epsilon_{p+1}' \\
        \vdots \\
        \epsilon_{T}'
    \end{pmatrix}}_{\mathcal{E}},
\end{equation}
and a vectorized version can be expressed as follows:
\begin{equation}
    \underbrace{\vecop(\mathcal{Y}_{Z})}_{Y} =
    \underbrace{(I_{d} \otimes \mathcal{X}_{Z})}_{\mathcal{Z}} 
    \underbrace{\vecop(B)}_{\beta} + 
    \underbrace{\vecop(\mathcal{E})}_{E},
    \label{al:VARlinear}
\end{equation}
where $Y \in \R^{Nd}$ with $N = T-p$, $Z \in \R^{Nd \times q}$ with $q = pd^2$, $\beta \in \R^{q}$, and $E \in \R^{Nd}$. The Yule--Walker estimator for $\beta$ can be constructed as the solution to the following objective:
\begin{equation}
    \label{eq:YW}
    \begin{gathered}
    \underset{\beta}{\text{argmin}} (-2 \beta' \widehat{\gamma} + 
    \beta' \widehat{\Gamma} \beta),
    \\
    \widehat{\beta} 
    = \widehat{\Gamma}^{-1} \widehat{\gamma},
    \end{gathered}
\end{equation}
with
\begin{equation} 
    \label{eq:Ghats}
    \begin{gathered}
        \widehat{\gamma}
        = \vecop( \widehat{\bm{\gamma}}_{Z} )
        = \vecop \bigl( \widehat{g}( \widehat{\bm{\gamma}}_{X} ) \bigr), \\
        \widehat{\Gamma} 
        = I_{d} \otimes \widehat{\boldsymbol{\Gamma}}_Z
        = I_{d} \otimes \widehat{g}( \widehat{\boldsymbol{\Gamma}}_X ),
    \end{gathered}
\end{equation}
where $\widehat{\bm{\gamma}}_{Z} = \bigl(\widehat{\Gamma}_{Z}(1)', \dots, \widehat{\Gamma}_{Z}(p)' \bigr)' $ and $\widehat{\boldsymbol{\Gamma}}_Z = \bigl(\widehat{\Gamma}_{Z}(r-s) \bigr)_{r,s = 1,\dots, p}$. That is, the Yule--Walker estimator for $\beta$ is specified in terms of the estimated autocovariances of the latent Gaussian series, which are obtained by applying the estimated inverse link function in \eqref{eq:inverse} to the estimated second-order properties of the observed discrete-valued process. The autocovariances of $\{\vect{X}_t\}$ can be estimated by setting $\widehat{\bm{\gamma}}_{X} = N^{-1} \mathcal{X}_{X}' \mathcal{Y}_{X}$ and $\widehat{\boldsymbol{\Gamma}}_X = N^{-1} \mathcal{X}_{X}' \mathcal{X}_{X}$, where $\mathcal{Y}_{X}$ and $\mathcal{X}_{X}$ are the discrete-valued analogues to the quantities specified in \eqref{eq:VARlinear}. 

\section{Analytic Standard Errors}

We now introduce analytic standard errors for the estimated latent Gaussian dynamics. We begin by stating the main result and then outline the derivation. Let $\widehat{\beta}$ denote the Yule--Walker estimator of the latent Gaussian dynamics in \eqref{eq:YW}, and let $\beta$ denote its population counterpart obtained by replacing the estimated latent Gaussian autocovariances in \eqref{eq:YW} with their population analogues. Then, the asymptotic distribution of the specified Yule--Walker estimator is multivariate normal:
\begin{equation}
    \label{eq:asymdist}
    \sqrt{T}(\widehat{\beta} - \beta) \overset{d}{\to} \mathcal{N}(0, \Sigma_Q),
\end{equation}
where $\Sigma_Q$ is the $q \times q$ limiting covariance matrix. Analytic standard errors for $\widehat{\beta}$ can therefore be obtained from the square roots of the diagonal entries of $\Sigma_Q/T$. Full details of the proof of asymptotic normality, together with the exact form of $\Sigma_Q$, are provided in the Appendix and Supplementary Materials.

To derive $\Sigma_Q$, note that the estimated Yule--Walker quantities are smooth functions of two components: the parameters of the marginal distributions and the autocovariances of $\{\vect{X}_t\}$. To make this dependence explicit, we can specify estimated latent Gaussian autocovariances in \eqref{eq:Ghats} as follows:
\begin{equation}
    \label{eq:fs}
    \begin{gathered}
        \widehat{\gamma}
        = h_1(\widehat{\theta}, \widehat{\bm{\Gamma}}^{p+1}_{X}), \\
        \widehat{\Gamma}
        = h_2(\widehat{\theta}, \widehat{\bm{\Gamma}}^{p+1}_{X}),
    \end{gathered}
\end{equation}
where $h_1$ and $h_2$ are fixed functions, $\widehat{\bm{\Gamma}}^{p+1}_{X} = \bigl( \widehat{\Gamma}_{X}(r-s) \bigr)_{r,s=1,\dots,p+1}$, and the corresponding population quantities are defined analogously. Note that the latent Gaussian autocovariances in \eqref{eq:Ghats} are equivalent to those in \eqref{eq:fs}. By the delta method, it therefore suffices to characterize the joint asymptotic distribution of $(\widehat{\gamma}, \widehat{\Gamma})$:
\begin{equation}
    \label{eq:delta1}
    \sqrt{T}
    \begin{pmatrix}
        \vecop(\widehat{\gamma}-\gamma) \\
        \vecop(\widehat{\Gamma}-\Gamma)
    \end{pmatrix}
    =
    \biggl(
        \sqrt{T}
        \bigl[
            \vecop\bigl(h_j(\widehat{\theta}, \widehat{\bm{\Gamma}}^{p+1}_{X})\bigr)
            -
            \vecop\bigl(h_j(\theta, \bm{\Gamma}^{p+1}_{X})\bigr)
        \bigr]
    \biggr)_{j=1,2}
    \overset{d}{\to} \mathcal{N}(0,\Sigma_R),
\end{equation}
where $\Sigma_R$ is the limiting covariance matrix of this joint distribution. A first-order Taylor expansion of $h_j(\widehat{\theta}, \widehat{\bm{\Gamma}}^{p+1}_{X})$ about $(\theta, \bm{\Gamma}^{p+1}_{X})$ yields that the left-hand side of \eqref{eq:delta1} can be approximated through 
%
\begin{equation}
    \label{eq:delta2}
    \left(
    \sqrt{T}
    \left[
    \frac{\partial \vecop \bigl( h_j (\theta, \bm{\Gamma}^{p+1}_{X}) \bigr)}
    {\partial \theta} \, (\widehat{\theta} - \theta)
    +
    \frac{\partial \vecop \bigl( h_j (\theta, \bm{\Gamma}^{p+1}_{X}) \bigr)}
    {\partial \vecop(\bm{\Gamma}^{p+1}_{X})}
    \vecop(\widehat{\bm{\Gamma}}^{p+1}_{X} - \bm{\Gamma}^{p+1}_{X})
    \right]
    \right)_{j=1,2}
    + o_{\Prob}(1).
\end{equation}
Hence, the limiting distribution in \eqref{eq:delta1} follows from the derivatives specified in \eqref{eq:delta2} and the joint asymptotic distribution of $(\widehat{\theta}, \widehat{\bm{\Gamma}}^{p+1}_{X})$:
\begin{equation}
    \label{eq:marg-cov dist}
    \sqrt{T}
    \begin{pmatrix}
        \widehat{\theta} - \theta \\
        \vecop(\widehat{\bm{\Gamma}}^{p+1}_{X} - \bm{\Gamma}^{p+1}_{X})
    \end{pmatrix}
    \overset{d}{\to} \mathcal{N}(0,\Sigma_Z),
\end{equation}
where $\Sigma_Z$ is the corresponding limiting covariance matrix.

The derivation of the asymptotic distribution of the estimated latent Gaussian dynamics in \eqref{eq:asymdist} therefore follows from two applications of the delta method. First, we derive the joint asymptotic distribution of $(\widehat{\gamma}, \widehat{\Gamma})$, with limiting covariance matrix $\Sigma_R$ determined by the derivatives specified in \eqref{eq:delta2} and the limiting covariance matrix $\Sigma_Z$ in \eqref{eq:marg-cov dist}. Then, because $\widehat{\beta}$ is a smooth function of $(\widehat{\gamma}, \widehat{\Gamma})$, defined by the Yule--Walker map, a second application of the delta method yields the asymptotic distribution and limiting covariance matrix $\Sigma_Q$ in \eqref{eq:asymdist}. In the following sections, we expand on the primary components of this derivation; that is, the limiting covariance matrix $\Sigma_Z$ and the derivatives in \eqref{eq:delta2}.

\subsection{Asymptotics of Marginal Parameters and Autocovariances}

We begin by characterizing the limiting covariance matrix \(\Sigma_Z\), which governs the joint asymptotic distribution of the estimators of the marginal parameters and the observed autocovariances. Because both sets of estimators are based on sample moments of the observed process \(\{X_t\}\), their joint asymptotic behavior can be obtained from a multivariate central limit theorem for a stacked vector collecting the sample quantities needed to estimate the marginal parameters and autocovariances. For simplicity, we present the argument for the \(p=1\) case. 

Let $\widehat{\theta}$ denote a method-of-moments estimator of the marginal parameter vector $\theta$. That is, $\widehat{\theta}$ is defined as the solution to the sample moment equations
\begin{equation}
    \label{eq:MoM-estimator}
    \frac1T\sum_{t=1}^T r(\vect{X}_t,\widehat\theta)=0.
\end{equation}
In the simplest case, components of $\widehat{\theta}$ may be specified as the sample average, as occurs for one-parameter marginal distributions whose parameter is identified by the mean, such as the Bernoulli and Poisson distributions. However, the same reasoning applies to other marginal distributions by replacing the sample mean with the appropriate sample moments identifying the marginal parameter vector $\theta$. Sample quantities of the observed process can similarly be used to estimate the required autocovariances. Indeed, when $p = 1$, $\bm{\Gamma}^{p+1}_{X}$ admits the following representation:
\begin{equation}
    \bm{\Gamma}^{2}_{X} =
    \begin{pmatrix}
        \Gamma_X(0) & \Gamma_X(-1) \\
        \Gamma_X(1) & \Gamma_X(0)
    \end{pmatrix},
\end{equation}
where, for a centered observed process $\{X_t\}$, estimates of $\Gamma_X(0)$, $\Gamma_X(1)$, and $\Gamma_X(-1)$ can be obtained from sample averages of $X_tX_t'$, $X_{t+1}X_t'$, and $X_{t-1}X_t'$, respectively.

The quantities defined above can then be combined to create the following vector process:
\begin{equation}
    \label{eq:Wt}
    W_t =
    \begin{pmatrix}
        -\,(J(\theta))^{-1} r(\vect{X}_t,\theta) \\
        \vecop
        \begin{pmatrix}
            \vect{X}_t \vect{X}^{'}_t \\
            \vect{X}_{t-1} \vect{X}^{'}_t
        \end{pmatrix} \\
        \vecop
        \begin{pmatrix}
            \vect{X}_{t+1} \vect{X}^{'}_{t} \\
            \vect{X}_t \vect{X}^{'}_t
        \end{pmatrix}
    \end{pmatrix},
\end{equation}
where $J(\theta)$ is the nonsingular Jacobian of the population moment equations with respect to $\theta$, which reduces to identity when $\theta$ is identified by the mean. Then, the following asymptotic representation holds:
\begin{equation}
    \label{eq:W-equal}
    \sqrt{T}
    \begin{pmatrix}
        \widehat{\theta}-\theta\\
        \vecop\bigl(\widehat{\bm{\Gamma}}^{p+1}_{X} -  \bm{\Gamma}^{p+1}_{X}\bigr)
    \end{pmatrix}
    =
    \frac{1}{\sqrt{T}}\sum_{t=1}^{T-p} W_t + o_{\Prob}(1).
\end{equation}
The joint asymptotic distribution of the estimated marginal parameters and autocovariances of $\{\vect{X_t}\}$ therefore follows from the asymptotic behavior of $W_t$:
\begin{equation}
    \label{eq:lrv}
    \frac{1}{\sqrt{T}}\sum_{t=1}^{T-p} W_t
    \xrightarrow{\operatorname{d}}
    \mathcal{N}(0,\Sigma_Z),
    \qquad
    \Sigma_Z=\sum_{k=-\infty}^{\infty}\Cov(W_0,W_k).
\end{equation}
An estimate of $\Sigma_Z$ can be obtained via estimation of the long-run variance of $\{W_t\}$, for which standard heteroskedasticity and autocorrelation consistent estimators, such as the Newey--West estimator, can be used.

\subsection{Derivatives with Respect to Autocovariances}

In addition to the limiting covariance matrix $\Sigma_Z$, the asymptotic representation in \eqref{eq:delta2} is characterized by derivatives of the fixed functions $h_1$ and $h_2$, which can be specified in terms of the derivatives of the inverse link function $g$. Continuing to emphasize dependence on $\theta$ by writing $g(\cdot,\theta)$ and $\ell(\cdot, \theta)$, the derivatives of the inverse link function $g$ with respect to the observed autocovariances can be computed element-wise:
\begin{equation}
    \label{eq:deriv-g}
    \frac{\partial}{\partial y} g_{ij}(y,\theta)
    =
    \frac{1}{
    \left.
    \frac{\partial}{\partial u} \ell_{ij}(u,\theta)
    \right|_{u = g_{ij}(y,\theta)}
    },
\end{equation}
where $y$ is an element of $\bm{\Gamma}^{p+1}_{X}$. By Proposition 2.1 in \textcite{Duker2024}, 
\begin{equation}
    \label{eq:deriv-ell}
    \frac{\partial}{\partial u} \ell_{ij}(u, \theta) =
    \frac{1}{2\pi \sqrt{1-u^2}} 
    \sum_{n_0,n_1=0}^\infty 
    \exp \left(-\frac{Q_{i,n_0}^2 + Q_{j,n_1}^2 - 2uQ_{i,n_0}Q_{j,n_1}}{2(1-u^2)}\right),
    \quad u\in(-1,1),
\end{equation}
with $Q_{i,n}$ defined as in \eqref{eq:representation_coef}. In the Bernoulli case, \eqref{eq:deriv-ell} admits a simplified representation:
\begin{equation}
    \label{eq:deriv-bern}
    \frac{\partial}{\partial u} \ell_{ij}(u,\theta) =
    \frac{1}{2\pi \sqrt{1-u^2}}\,
    \exp \left(-\frac{q_i^2+q_j^2-2uq_iq_j}{2(1-u^2)}\right),
\end{equation}
where $q_i = \Phi^{-1}(1 - p_i)$. Thus, \eqref{eq:deriv-g} can be computed explicitly when the marginal distributions of $\{X_t\}$ are Bernoulli. When the marginal distributions are not Bernoulli, \eqref{eq:deriv-g} can be calculated numerically. 

\subsection{Derivatives with Respect to Marginal Parameters}

The derivatives with respect to the marginal parameters in \eqref{eq:delta2} can similarly be specified in terms of $g$ and $\ell$. First, we note that, by definition,
\begin{equation}
    \label{eq:chain-g-ell}
    \frac{\partial}{\partial\theta}\, 
    \vecop \biggl( g \bigl( \ell(\bm{\Gamma}^{p+1}_{Z},\theta),\theta \bigr) \biggr)
    = \frac{\partial}{\partial\theta}\,
    \vecop (\bm{\Gamma}^{p+1}_{Z})
    = 0,
\end{equation}
where $\bm{\Gamma}^{p+1}_{Z} = \bigl( \Gamma_{Z}(r-s) \bigr)_{r,s=1,\dots,p+1}$. Then, for $U = \ell(\bm{\Gamma}^{p+1}_{Z},\theta) = \bm{\Gamma}_X^{p+1}$, an application of the chain rule yields
\begin{equation}
    \label{eq:deriv-theta}
    \frac{\partial}{\partial\theta}\,\vecop \bigl(g(U,\theta)\bigr)
    =
    -
    \left.\frac{\partial}{\partial\vecop(U)}\,\vecop \bigl(g(U,\theta)\bigr)\right|_{U=\bm{\Gamma}_X^{p+1}}
    \frac{\partial}{\partial\theta}\, 
    \vecop \bigl(\ell(\bm{\Gamma}_Z^{p+1},\theta)\bigr).
\end{equation}
As $g(\cdot,\theta)$ is applied element-wise, the Jacobian with respect to $\vecop(U)$ is a diagonal matrix with elements computed according to \eqref{eq:deriv-g}. The second Jacobian encodes the partial derivatives of the link function $\ell$ with respect to $\theta$. Specifically, for $a\in\{1,\dots,d\}$, let $M^a=(M^a(r-s))_{r,s=1,\dots,p+1}$ be a $(p+1)\times(p+1)$ block matrix, where each $d \times d$ block $M^a(r)=(M^a_{ij}(r))_{i,j=1,\dots,d}$ contains the element-wise derivatives of $\ell(\Gamma_Z(r-s),\theta)$ with respect to $\theta_a$. Then,
\begin{equation}
    \frac{\partial}{\partial\theta}\,\vecop \bigl(\ell(\bm{\Gamma}_Z^{p+1},\theta)\bigr)
    =
    \begin{pmatrix}
        \vecop(M^1) & \cdots & \vecop(M^d)
    \end{pmatrix},
\end{equation}
where each column represents the derivatives of $\bm{\Gamma}_X^{p+1}$ with respect to an element of $\theta$. When the marginal distributions of $\{X_t\}$ are Bernoulli, $M^a(r)$ admits the following representation:
\begin{equation}
    \label{eq:Msij-def}
    M^a_{ij}(r)
    =
    \begin{cases}
        0, & a\notin\{i,j\},\\[0.4em]
        \sqrt{2\pi}\,e^{q_i^2/2}\displaystyle\sum_{k=1}^\infty \frac{1}{k!}\,c_{i,k+1}c_{j,k}\,
        \bigl(g_{ij}(\Gamma_{X,ij}(r),\theta)\bigr)^k,
        & a=i,\ i\neq j,\\[0.8em]
        \sqrt{2\pi}\,e^{q_j^2/2}\displaystyle\sum_{k=1}^\infty \frac{1}{k!}\,c_{i,k}c_{j,k+1}\,
        \bigl(g_{ij}(\Gamma_{X,ij}(r),\theta)\bigr)^k,
        & a=j,\ i\neq j,\\[0.8em]
        2\sqrt{2\pi}\,e^{q_i^2/2}\displaystyle\sum_{k=1}^\infty \frac{1}{k!}\,c_{i,k+1}c_{i,k}\,
        \bigl(g_{ii}(\Gamma_{X,ii}(r),\theta)\bigr)^k,
        & a=i=j.
    \end{cases}
\end{equation}
When the marginal distributions are not Bernoulli, these derivatives can be computed numerically.

\section{Simulation Study}

To evaluate the performance of the estimated dynamics and asymptotic standard errors, we conducted a Monte Carlo simulation study. Performance was assessed across a range of factors, including number of variables $ d \in \{3,5\}$, time series length $T \in \{50,100,200,500\}$, marginal distributions (Bernoulli, Poisson, mixed), and magnitude of the marginal parameters (small, medium, large). The levels of the marginal distribution factor describe the composition of the observed discrete-valued process. The Bernoulli and Poisson conditions, for example, had marginal distributions that were exclusively Bernoulli and Poisson, respectively. Alternatively, the mixed condition consisted of a combination of Bernoulli, Poisson, and standard Gaussian marginal distributions. When $d = 5$, the marginal distributions in the mixed conditions consisted of two Bernoulli, two Poisson, and one standard Gaussian. The magnitudes of the marginal parameters were specified as follows. In the small magnitude condition, the Bernoulli and Poisson parameters were set to $p = 0.3$ and $\lambda = 1$, respectively. In the medium magnitude condition, the parameters were specified as $p = 0.5$ and $\lambda = 5$. In the large magnitude condition, the Bernoulli marginal parameter was $p = 0.7$, and the Poisson marginal parameter was $\lambda = 10$. Note that the parameters of the standard Gaussian marginal distribution were $\mu = 0$ and $\sigma^2 = 1$ across all conditions. For each condition, 500 replications were generated, yielding a fully factorial design with 72 conditions  $(2 \times 4 \times 3 \times 3)$ and 36,000 $(72 \times 500)$ unique datasets. All R code and data required to reproduce the simulation study are publicly available at \url{https://anonymous.4open.science/r/LatentGaussianVAR-8B2A/}.

\subsection{Data Generation and Model Estimation}

All data were generated according to the following procedure. First, the latent Gaussian series, $\{\vect{Z}_t\}$, was specified as a VAR(1) process. Each element of the $d \times d$ transition matrix, $\Phi$, was drawn from a $\mathcal{U}(-0.4,0.4)$ distribution until a stationary solution was obtained. This range was selected to yield dynamics typical in the social, behavioral, and health sciences, with both positive and negative autocovariances represented. Note that the generated transition matrices did not vary by replication. That is, the data generating process was consistent within each condition. This enabled comparisons between the analytic standard errors and empirical standard errors, which are defined below. The covariance matrix of the white noise innovation process, $\Sigma_\epsilon$, was specified as identity. Unit variance of the latent process was induced by setting $\Phi^* =  S^{-1} \Phi S$ and $\Sigma_\epsilon^* = S^{-1} \Sigma_\epsilon S^{-1}$, where $S = \diag (\Gamma_Z(0))^{\frac{1}{2}}$. The discrete-valued process, $\{\vect{X}_t\}$, was then obtained by applying the relation in \eqref{eq:marginal} to the generated latent Gaussian VAR(1) process.

Estimation of model parameters and analytic standard errors proceeded in the manner detailed previously. The derivatives of the inverse link function, $g$, in \eqref{eq:deriv-g} and \eqref{eq:deriv-theta} were approximated using the \texttt{numDeriv} R package \parencite{numderiv}. The limiting covariance matrix of the joint asymptotic distribution of the marginal parameters and observed autocovariances in \eqref{eq:marg-cov dist} was estimated using the \texttt{cointReg} R package \parencite{cointreg}. To contextualize performance, we compared estimates with a ``naive'' benchmark obtained by fitting a canonical Gaussian VAR(1) model to the discrete-valued series. This approach is common in other domains \parencite[e.g.,][]{Rhemtulla2012}, and therefore provides a plausible alternative. Data generation and model estimation procedures were implemented in the \texttt{timecop} R package. The canonical Gaussian VAR(1) models were estimated using the \texttt{vars} R package \parencite{vars}.

\subsection{Outcome Measures}

The performance of the estimated dynamics and asymptotic standard errors was assessed in the context of several metrics, including absolute bias, relative bias, root mean square error (RMSE), and coverage rate. Absolute bias, relative bias, and RMSE were defined as follows:
\begin{align*}
    \text{Absolute Bias}
    &= \frac{1}{d^2} \sum_{i=1}^d \sum_{j=1}^d 
    \left| \widehat{\phi}_{ij} - \phi_{ij} \right|, 
    \\[0.8em]
    \text{Relative Bias}
    &= \operatorname{median}_{1\le i\le d,\;1\le j\le d}
    \left( \frac{\widehat{\phi}_{ij} - \phi_{ij}}{\phi_{ij}} \right),
    \\[0.8em]
    \text{RMSE}
    &= \sqrt{ \frac{1}{d^2} \sum_{i=1}^d \sum_{j=1}^d 
    \big(\widehat{\phi}_{ij} - \phi_{ij}\big)^2 },
\end{align*}
where $\widehat{\phi}_{ij}$ and $\phi_{ij}$ were the estimated and true quantities of interest, respectively. For the estimated dynamics, these quantities corresponded to entry $(i,j)$ of the estimated and true transition matrices, respectively, where the true transition matrix was defined as $\boldsymbol{\Phi}^*$ above. For the estimated asymptotic standard errors, $\widehat{\phi}_{ij}$ was entry $(i,j)$ of the estimated matrix of standard errors. The corresponding true element was the empirical standard error, which was defined as the standard deviation of entry $(i,j)$ of the estimated transition matrix across replications. Note that we considered the median value for relative bias, as this metric can produce extreme values when the true parameters are small in magnitude. Coverage rate was defined as the proportion of 95\% confidence intervals that included the true parameter across replications.

\subsection{Results}

\paragraph{Nonstationary Solutions.} Prior to reporting results of the simulation study, we first make note of conditions for which the latent Gaussian approach produced nonstationary solutions. As described previously, the Yule--Walker estimates of the latent dynamics are obtained via the second-order properties of the latent Gaussian process, which are estimated through spline-based interpolation of the inverse link function in \eqref{eq:inverse}. Importantly, this procedure does not guarantee that $\widehat{\Gamma}_{Z}(0)$ is non-negative definite \parencite{Kim2025}. When this occurs, or when the estimated latent covariance matrix is ill-conditioned, the resulting Yule--Walker estimates can imply nonstationary transition matrices, as the estimated latent second-order properties do not correspond to a valid stationary covariance structure. In the current simulation study, this was confined almost exclusively to simulation conditions characterized by a combination of short time series length, large number of component series, and marginal parameters near the boundary of the parameter space, indicating that these solutions are a product of processes that convey limited information relative to the number of parameters being estimated. To address these cases, we explored two potential remedies. First, nonstationary solutions were discarded and simulation results were assessed with respect to the remaining replications. Second, we employed the approach proposed by Higham (\citeyear{Higham2002}), which computes the nearest positive semidefinite matrix given a non-positive definite matrix, where distance is defined with respect to the Frobenius norm. This positive semidefinite matrix was then used to compute the Yule--Walker estimates of the latent Gaussian dynamics. Results of the current simulation study did not appreciably differ between the two approaches. We therefore only report the results for which the inadmissible solutions were discarded. 

\begin{figure}
    \centering
    \includegraphics[width=1\linewidth]{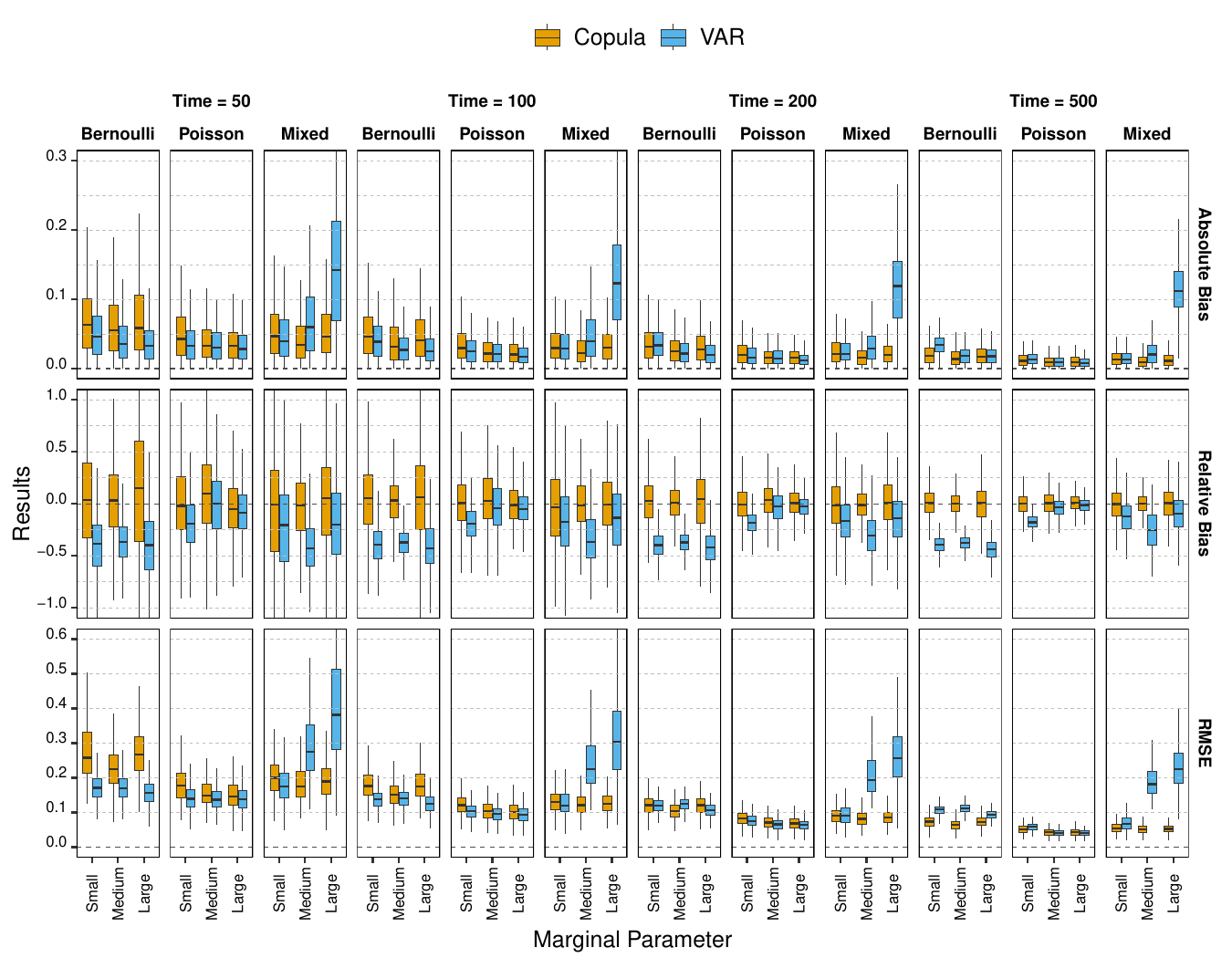}
    \caption{Simulation results for point estimates when $d = 3$.}
    \label{fig:est3}
\end{figure}

\begin{figure}
    \centering
    \includegraphics[width=1\linewidth]{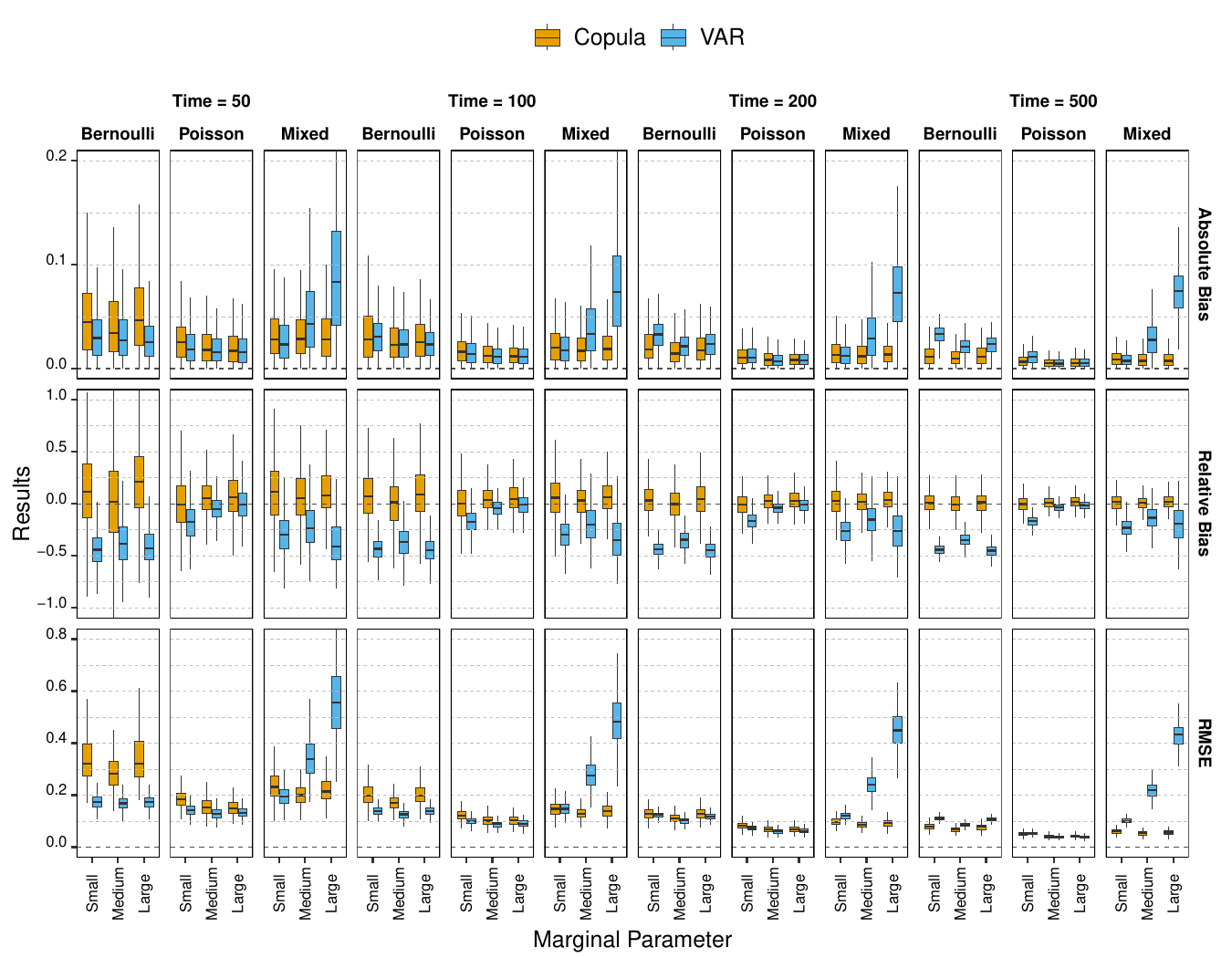}
    \caption{Simulation results for point estimates when $d = 5$.}
    \label{fig:est5}
\end{figure}

\paragraph{Point Estimates.} Point estimate absolute bias, relative bias, and RMSE for each condition can be found in Table \ref{tab:est}, and these results are visualized in Figures \ref{fig:est3} and \ref{fig:est5}. Across all conditions, absolute bias decreased as time series length increased for both the latent Gaussian and canonical VAR models. In the mixed-distribution conditions, point estimates for the latent Gaussian model had consistently lower absolute and relative bias than the corresponding canonical VAR estimates. In the three-variable conditions, for example, absolute bias ranged from 0.012 to 0.055 for the latent Gaussian model and from 0.016 to 0.150 for the canonical VAR. In the Bernoulli and Poisson conditions, differences in absolute bias between the latent Gaussian and canonical VAR approaches were generally smaller. When time series length was 50 or 100, the canonical VAR often produced point estimates with lower absolute bias; however, this pattern was reduced or reversed at longer time series lengths. In the Poisson case, absolute bias of the latent Gaussian dynamics generally mirrored that of the canonical VAR when time series length was 200 or 500. In the five-variable Bernoulli conditions, the latent Gaussian dynamics exhibited smaller absolute bias than the canonical VAR estimates when time series length was 200 or 500.

The relative bias of the latent Gaussian point estimates similarly decreased as time series length increased, whereas the canonical VAR dynamics were characterized by a more stable pattern. In the Bernoulli three-variable conditions, for example, the relative bias of the canonical VAR effects ranged from $-0.434$ to $-0.366$, whereas the relative bias of the latent Gaussian estimates ranged from 0.002 to 0.131. In the mixed marginal distribution conditions, the latent Gaussian point estimates had lower relative bias than the canonical VAR effects across all combinations of time series length, number of component series, and magnitude of marginal parameters. Relative bias of the latent Gaussian point estimates was largest in the five-variable Bernoulli conditions when the marginal parameters were either small or large, particularly when time series length was small. Finally, RMSE of point estimates for both the latent Gaussian and canonical VAR dynamics generally mirrored the patterns observed for absolute bias, but clarified differences in variability. In several mixed distribution cases, the canonical VAR effects displayed similar absolute bias but larger RMSE than the latent Gaussian estimates, indicating greater variability of the canonical VAR point estimates. Similarly, some five-variable Bernoulli conditions exhibited lower absolute bias than the corresponding three-variable conditions, but higher RMSE, suggesting increased variability for both models.

\begin{landscape}
\begin{table}[ht]
\caption{Point Estimate Absolute Bias, Relative Bias, and RMSE} 
\label{tab:est}
\centering
{
\renewcommand{\arraystretch}{1.5}
\scalebox{0.62}{
\begin{tabular}{rrrrrrrrrrrrrrrrrrrrr}
\toprule
&&& \multicolumn{18}{c}{Measure}\\

\cmidrule(lr){4-21}

&&& \multicolumn{6}{c}{Absolute Bias} & \multicolumn{6}{c}{Relative Bias} & \multicolumn{6}{c}{RMSE} \\

\cmidrule(lr){4-9} \cmidrule(lr){10-15} \cmidrule(lr){16-21}

&&& \multicolumn{6}{c}{Marginal Distribution} & \multicolumn{6}{c}{Marginal Distribution} & \multicolumn{6}{c}{Marginal Distribution} \\

\cmidrule(lr){4-9} \cmidrule(lr){10-15} \cmidrule(lr){16-21}

&&& \multicolumn{2}{c}{Bernoulli} & \multicolumn{2}{c}{Poisson} & \multicolumn{2}{c}{Mixed} & \multicolumn{2}{c}{Bernoulli} & \multicolumn{2}{c}{Poisson} & \multicolumn{2}{c}{Mixed} & \multicolumn{2}{c}{Bernoulli} & \multicolumn{2}{c}{Poisson} & \multicolumn{2}{c}{Mixed} \\

\cmidrule(lr){4-5}\cmidrule(lr){6-7}\cmidrule(lr){8-9}\cmidrule(lr){10-11}\cmidrule(lr){12-13}\cmidrule(lr){14-15}\cmidrule(lr){16-17}\cmidrule(lr){18-19}\cmidrule(lr){20-21}

\multicolumn{1}{c}{Variables} &\multicolumn{1}{c}{Time} &\multicolumn{1}{c}{Magnitude} & \multicolumn{1}{c}{Copula} & \multicolumn{1}{c}{VAR} & \multicolumn{1}{c}{Copula} & \multicolumn{1}{c}{VAR} & \multicolumn{1}{c}{Copula} & \multicolumn{1}{c}{VAR} & \multicolumn{1}{c}{Copula} & \multicolumn{1}{c}{VAR} & \multicolumn{1}{c}{Copula} & \multicolumn{1}{c}{VAR} & \multicolumn{1}{c}{Copula} & \multicolumn{1}{c}{VAR} & \multicolumn{1}{c}{Copula} & \multicolumn{1}{c}{VAR} & \multicolumn{1}{c}{Copula} & \multicolumn{1}{c}{VAR} & \multicolumn{1}{c}{Copula} & \multicolumn{1}{c}{VAR}\\
 \midrule
3 & 50 & Small & 0.078 & 0.051 & 0.050 & 0.037 & 0.055 & 0.049 & 0.042 & -0.395 & 0.020 & -0.187 & -0.045 & -0.234 & 0.285 & 0.173 & 0.185 & 0.143 & 0.206 & 0.181 \\ 
  3 & 50 & Medium & 0.065 & 0.042 & 0.039 & 0.036 & 0.042 & 0.072 & 0.046 & -0.373 & 0.097 & -0.008 & -0.018 & -0.398 & 0.232 & 0.173 & 0.157 & 0.140 & 0.189 & 0.299 \\ 
  3 & 50 & Large & 0.073 & 0.039 & 0.037 & 0.035 & 0.055 & 0.150 & 0.131 & -0.408 & -0.038 & -0.087 & 0.041 & -0.177 & 0.275 & 0.158 & 0.156 & 0.139 & 0.195 & 0.413 \\ 
  3 & 100 & Small & 0.052 & 0.041 & 0.036 & 0.028 & 0.036 & 0.034 & 0.049 & -0.395 & 0.018 & -0.189 & -0.038 & -0.180 & 0.181 & 0.138 & 0.124 & 0.103 & 0.134 & 0.129 \\ 
  3 & 100 & Medium & 0.041 & 0.031 & 0.026 & 0.025 & 0.028 & 0.049 & 0.028 & -0.372 & 0.054 & -0.029 & -0.006 & -0.338 & 0.154 & 0.141 & 0.105 & 0.096 & 0.126 & 0.242 \\ 
  3 & 100 & Large & 0.050 & 0.030 & 0.025 & 0.022 & 0.035 & 0.125 & 0.070 & -0.416 & 0.008 & -0.043 & 0.001 & -0.153 & 0.183 & 0.126 & 0.104 & 0.094 & 0.128 & 0.316 \\ 
  3 & 200 & Small & 0.037 & 0.037 & 0.024 & 0.019 & 0.026 & 0.026 & 0.030 & -0.396 & 0.012 & -0.178 & -0.006 & -0.161 & 0.123 & 0.120 & 0.085 & 0.077 & 0.090 & 0.096 \\ 
  3 & 200 & Medium & 0.029 & 0.025 & 0.018 & 0.018 & 0.018 & 0.034 & 0.021 & -0.366 & 0.035 & -0.029 & -0.006 & -0.297 & 0.105 & 0.124 & 0.072 & 0.066 & 0.083 & 0.209 \\ 
  3 & 200 & Large & 0.033 & 0.023 & 0.018 & 0.014 & 0.023 & 0.116 & 0.027 & -0.423 & 0.015 & -0.027 & 0.009 & -0.141 & 0.121 & 0.106 & 0.071 & 0.065 & 0.088 & 0.264 \\ 
  3 & 500 & Small & 0.021 & 0.035 & 0.014 & 0.015 & 0.016 & 0.016 & 0.015 & -0.394 & 0.003 & -0.175 & 0.001 & -0.131 & 0.074 & 0.108 & 0.052 & 0.059 & 0.057 & 0.070 \\ 
  3 & 500 & Medium & 0.018 & 0.019 & 0.011 & 0.011 & 0.012 & 0.024 & 0.007 & -0.371 & 0.014 & -0.031 & 0.008 & -0.251 & 0.065 & 0.112 & 0.044 & 0.041 & 0.052 & 0.190 \\ 
  3 & 500 & Large & 0.021 & 0.019 & 0.011 & 0.010 & 0.014 & 0.114 & 0.002 & -0.434 & 0.013 & -0.014 & 0.004 & -0.099 & 0.074 & 0.094 & 0.044 & 0.040 & 0.054 & 0.230 \\ 
  5 & 50 & Small & 0.053 & 0.033 & 0.029 & 0.022 & 0.035 & 0.028 & 0.133 & -0.434 & 0.006 & -0.178 & 0.127 & -0.296 & 0.358 & 0.175 & 0.192 & 0.144 & 0.246 & 0.198 \\ 
  5 & 50 & Medium & 0.046 & 0.032 & 0.023 & 0.019 & 0.034 & 0.052 & 0.027 & -0.385 & 0.080 & -0.047 & 0.078 & -0.210 & 0.297 & 0.169 & 0.164 & 0.130 & 0.208 & 0.346 \\ 
  5 & 50 & Large & 0.058 & 0.029 & 0.022 & 0.019 & 0.034 & 0.093 & 0.212 & -0.416 & 0.089 & -0.006 & 0.113 & -0.366 & 0.356 & 0.173 & 0.158 & 0.133 & 0.234 & 0.568 \\ 
  5 & 100 & Small & 0.035 & 0.032 & 0.019 & 0.017 & 0.024 & 0.021 & 0.087 & -0.440 & 0.008 & -0.170 & 0.071 & -0.289 & 0.208 & 0.141 & 0.125 & 0.101 & 0.152 & 0.150 \\ 
  5 & 100 & Medium & 0.028 & 0.026 & 0.015 & 0.013 & 0.021 & 0.040 & 0.017 & -0.365 & 0.055 & -0.039 & 0.035 & -0.191 & 0.173 & 0.126 & 0.107 & 0.089 & 0.130 & 0.281 \\ 
  5 & 100 & Large & 0.032 & 0.025 & 0.014 & 0.013 & 0.022 & 0.079 & 0.098 & -0.447 & 0.061 & -0.013 & 0.069 & -0.327 & 0.206 & 0.139 & 0.104 & 0.091 & 0.142 & 0.491 \\ 
  5 & 200 & Small & 0.023 & 0.033 & 0.013 & 0.013 & 0.016 & 0.014 & 0.042 & -0.440 & -0.003 & -0.163 & 0.032 & -0.262 & 0.130 & 0.124 & 0.084 & 0.074 & 0.099 & 0.121 \\ 
  5 & 200 & Medium & 0.018 & 0.023 & 0.010 & 0.009 & 0.015 & 0.033 & -0.001 & -0.350 & 0.033 & -0.039 & 0.024 & -0.148 & 0.112 & 0.103 & 0.072 & 0.062 & 0.088 & 0.243 \\ 
  5 & 200 & Large & 0.021 & 0.024 & 0.010 & 0.009 & 0.016 & 0.073 & 0.051 & -0.442 & 0.041 & -0.010 & 0.044 & -0.261 & 0.131 & 0.119 & 0.071 & 0.063 & 0.094 & 0.452 \\ 
  5 & 500 & Small & 0.013 & 0.033 & 0.008 & 0.012 & 0.010 & 0.009 & 0.014 & -0.440 & 0.000 & -0.163 & 0.016 & -0.230 & 0.080 & 0.113 & 0.051 & 0.054 & 0.061 & 0.102 \\ 
  5 & 500 & Medium & 0.012 & 0.021 & 0.006 & 0.006 & 0.009 & 0.028 & 0.002 & -0.344 & 0.015 & -0.036 & 0.011 & -0.134 & 0.068 & 0.087 & 0.043 & 0.039 & 0.054 & 0.220 \\ 
  5 & 500 & Large & 0.014 & 0.023 & 0.006 & 0.006 & 0.009 & 0.075 & 0.018 & -0.447 & 0.024 & -0.012 & 0.022 & -0.199 & 0.078 & 0.107 & 0.043 & 0.039 & 0.056 & 0.431 \\ 
  \bottomrule
  \end{tabular}
}
}
\end{table}
\end{landscape}

\begin{figure}
    \centering
    \includegraphics[width=1\linewidth]{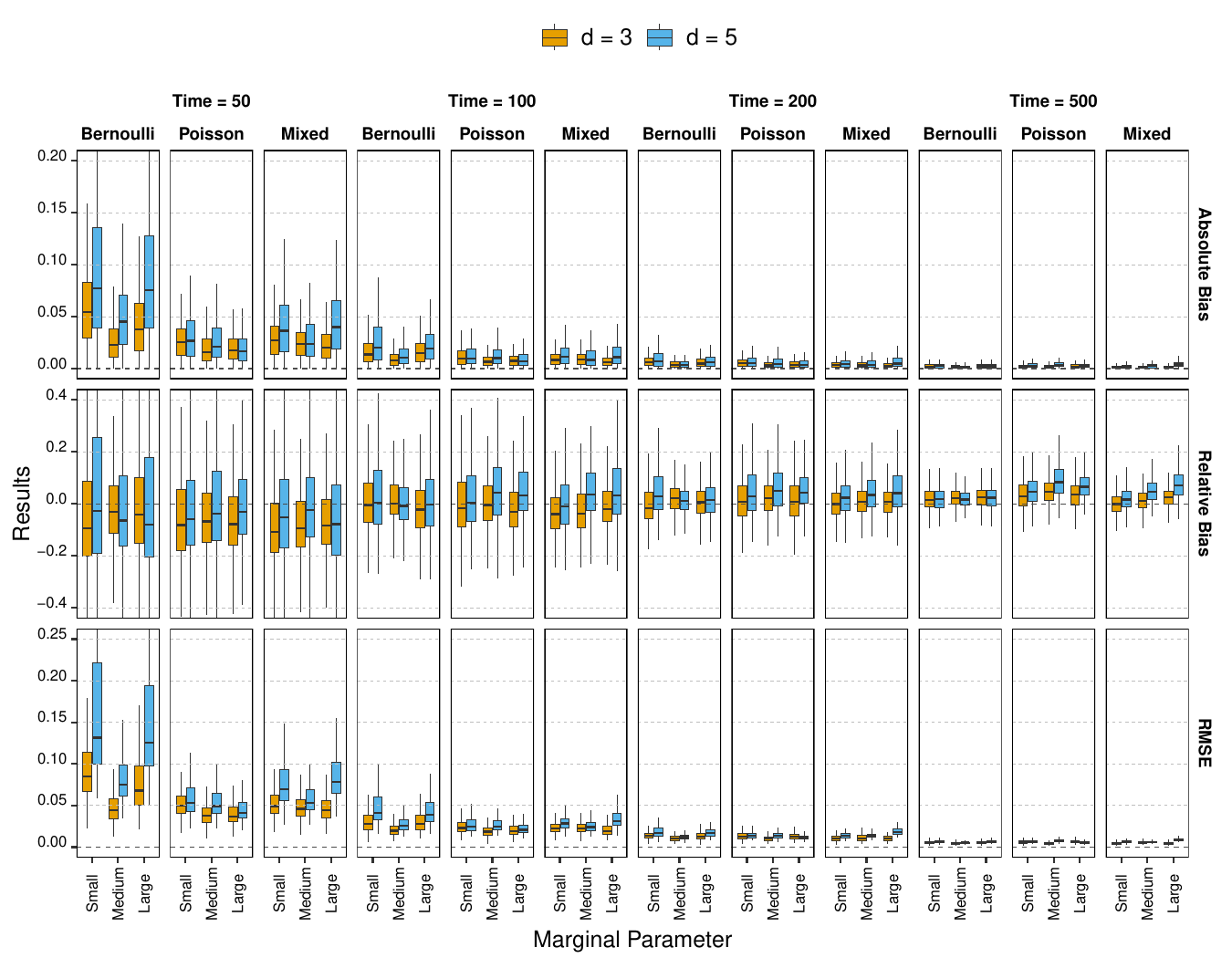}
    \caption{Simulation results for standard errors.}
    \label{fig:ses}
\end{figure}

\paragraph{Standard Errors.} Absolute bias, relative bias, and RMSE of estimated standard errors are shown in Table \ref{tab:se} and Figure \ref{fig:ses}. The ``true'' standard errors used to compute these metrics were defined as the standard deviation of the point estimates within a given condition. Because these empirical standard errors are estimator-specific, there is no common reference value that can be used to compare the estimated latent Gaussian standard errors with those of the canonical VAR. We therefore refrain from making such comparisons and report only the performance of the latent Gaussian estimated standard errors.

Across all conditions, absolute bias of the estimated latent Gaussian standard errors decreased as time series length increased. Absolute bias was smallest for the Poisson and mixed distribution conditions, and differences between the two were generally small in magnitude. In the three-variable conditions, for example, absolute bias ranged from 0.002 to 0.032 for mixed marginal distributions and from 0.003 to 0.033 for the Poisson distributions. Absolute bias was generally largest for the Bernoulli distributed processes; however, distributional differences were reduced when the magnitudes of the marginal parameters were medium. For example, absolute bias for the three-variable Bernoulli case was more than twice that of the mixed distribution condition when time series length was 100 and marginal parameters were small in magnitude, whereas no difference was observed when the magnitude of the marginal parameters was medium.

Relative bias showed a somewhat different pattern. The Poisson marginal distribution conditions frequently exhibited a greater degree of relative bias than the Bernoulli or mixed distribution cases, despite having small absolute bias. Relative bias was positive across most conditions, indicating that estimated standard errors were, on average, too large. Notably, relative bias was generally negative in the three-variable mixed distribution conditions when time series length was 50 or 100, indicating that standard errors were underestimated in these conditions. Finally, RMSE of the estimated standard errors generally mirrored the patterns observed for absolute bias. The Bernoulli marginal distribution conditions produced standard error estimates with larger RMSE values than the Poisson and mixed distribution cases, whereas differences between the Poisson and mixed distribution conditions were small and disappeared when time series length was 500. These results contrast those observed for point estimates, where differences between marginal distribution conditions in both absolute bias and RMSE were evident even at large time series lengths.

\clearpage
\begin{table}[ht]
\caption{Standard Error Absolute Bias, Relative Bias, and RMSE} 
\label{tab:se}
\centering
{
\renewcommand{\arraystretch}{1.5}
\scalebox{0.72}{
\begin{tabular}{rrrrrrrrrrrr}
\toprule

&&& \multicolumn{9}{c}{Measure}\\

\cmidrule(lr){4-12} 

&&& \multicolumn{3}{c}{Absolute Bias} & \multicolumn{3}{c}{Relative Bias} & \multicolumn{3}{c}{RMSE} \\

\cmidrule(lr){4-6} \cmidrule(lr){7-9} \cmidrule(lr){10-12} 

&&& \multicolumn{3}{c}{Marginal Distribution} & \multicolumn{3}{c}{Marginal Distribution} & \multicolumn{3}{c}{Marginal Distribution} \\

\cmidrule(lr){4-6} \cmidrule(lr){7-9} \cmidrule(lr){10-12} 

\multicolumn{1}{c}{Variables} &\multicolumn{1}{c}{Time} &\multicolumn{1}{c}{Magnitude} & \multicolumn{1}{c}{Bernoulli} & \multicolumn{1}{c}{Poisson} & \multicolumn{1}{c}{Mixed} & \multicolumn{1}{c}{Bernoulli} & \multicolumn{1}{c}{Poisson} & \multicolumn{1}{c}{Mixed} & \multicolumn{1}{c}{Bernoulli} & \multicolumn{1}{c}{Poisson} & \multicolumn{1}{c}{Mixed} \\
 \midrule
3 & 50 & Small & 0.070 & 0.033 & 0.032 & -0.012 & -0.019 & -0.065 & 0.111 & 0.059 & 0.056 \\ 
  3 & 50 & Medium & 0.033 & 0.021 & 0.030 & 0.005 & -0.033 & -0.036 & 0.055 & 0.042 & 0.053 \\ 
  3 & 50 & Large & 0.057 & 0.026 & 0.025 & 0.014 & -0.018 & -0.055 & 0.095 & 0.048 & 0.049 \\ 
  3 & 100 & Small & 0.023 & 0.014 & 0.011 & 0.036 & 0.027 & -0.018 & 0.040 & 0.027 & 0.025 \\ 
  3 & 100 & Medium & 0.011 & 0.009 & 0.011 & 0.023 & 0.013 & -0.013 & 0.022 & 0.020 & 0.025 \\ 
  3 & 100 & Large & 0.021 & 0.010 & 0.008 & 0.011 & -0.001 & 0.006 & 0.037 & 0.023 & 0.022 \\ 
  3 & 200 & Small & 0.008 & 0.006 & 0.004 & 0.004 & 0.027 & 0.009 & 0.016 & 0.014 & 0.011 \\ 
  3 & 200 & Medium & 0.005 & 0.005 & 0.004 & 0.029 & 0.030 & 0.016 & 0.011 & 0.011 & 0.012 \\ 
  3 & 200 & Large & 0.007 & 0.005 & 0.004 & 0.011 & 0.023 & 0.014 & 0.015 & 0.014 & 0.011 \\ 
  3 & 500 & Small & 0.003 & 0.003 & 0.002 & 0.022 & 0.036 & 0.005 & 0.006 & 0.006 & 0.005 \\ 
  3 & 500 & Medium & 0.002 & 0.003 & 0.002 & 0.026 & 0.054 & 0.016 & 0.005 & 0.005 & 0.006 \\ 
  3 & 500 & Large & 0.003 & 0.003 & 0.002 & 0.030 & 0.040 & 0.031 & 0.006 & 0.007 & 0.005 \\ 
  5 & 50 & Small & 0.142 & 0.052 & 0.052 & 0.128 & 0.047 & 0.012 & 0.224 & 0.087 & 0.091 \\ 
  5 & 50 & Medium & 0.064 & 0.035 & 0.036 & 0.031 & 0.041 & 0.030 & 0.101 & 0.066 & 0.068 \\ 
  5 & 50 & Large & 0.161 & 0.024 & 0.065 & 0.108 & 0.033 & 0.027 & 0.245 & 0.050 & 0.114 \\ 
  5 & 100 & Small & 0.036 & 0.018 & 0.016 & 0.080 & 0.062 & 0.028 & 0.061 & 0.033 & 0.034 \\ 
  5 & 100 & Medium & 0.016 & 0.016 & 0.013 & 0.032 & 0.081 & 0.066 & 0.032 & 0.031 & 0.028 \\ 
  5 & 100 & Large & 0.029 & 0.011 & 0.018 & 0.039 & 0.087 & 0.088 & 0.053 & 0.024 & 0.038 \\ 
  5 & 200 & Small & 0.012 & 0.008 & 0.007 & 0.065 & 0.062 & 0.040 & 0.021 & 0.016 & 0.015 \\ 
  5 & 200 & Medium & 0.005 & 0.008 & 0.006 & 0.021 & 0.080 & 0.057 & 0.013 & 0.016 & 0.016 \\ 
  5 & 200 & Large & 0.009 & 0.006 & 0.008 & 0.040 & 0.061 & 0.074 & 0.019 & 0.012 & 0.020 \\ 
  5 & 500 & Small & 0.004 & 0.004 & 0.003 & 0.032 & 0.061 & 0.031 & 0.008 & 0.007 & 0.007 \\ 
  5 & 500 & Medium & 0.002 & 0.005 & 0.004 & 0.023 & 0.110 & 0.065 & 0.006 & 0.008 & 0.007 \\ 
  5 & 500 & Large & 0.003 & 0.004 & 0.005 & 0.031 & 0.095 & 0.086 & 0.008 & 0.007 & 0.010 \\ 
  \bottomrule
  \end{tabular}
}
}
\end{table}

\begin{figure}[h]
    \centering
    \includegraphics[width=0.84\linewidth]{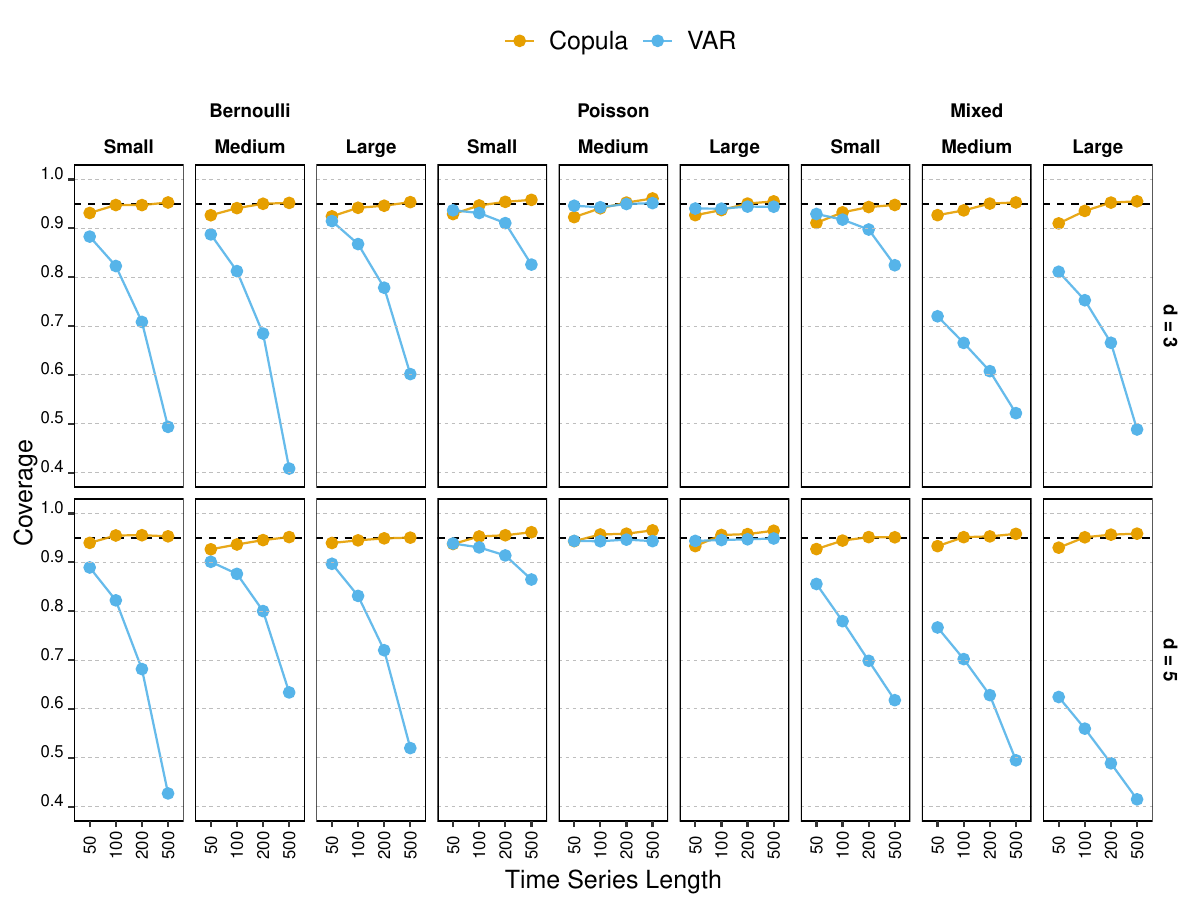}
    \caption{Empirical coverage rates for 95\% confidence intervals.}
    \label{fig:ci}
\end{figure}

\paragraph{Confidence Interval Coverage.} Confidence interval coverage rates are shown in Table \ref{tab:ci} and Figure \ref{fig:ci}. Coverage rates for the latent Gaussian point estimates were close to the nominal 0.95 level across all conditions. In the Bernoulli distribution conditions, coverage rates improved as time series length increased, with rates ranging from 0.945 to 0.955 for lengths of 200 and 500. A similar pattern was observed for the mixed distribution conditions, though slight over-coverage was detected in the five-variable cases when time series length was 500. In the Poisson distribution conditions, coverage rates were also close to the nominal level, though some over-coverage was observed at longer time series lengths.

In contrast, confidence interval coverage rates for the canonical VAR estimates frequently failed to reach the nominal 0.95 level. In the Bernoulli and mixed distribution conditions, coverage rates ranged from 0.408 to 0.915 and 0.415 to 0.929, respectively, with the degree of under-coverage becoming more pronounced at longer time series lengths. For example, coverage reached 0.915 in the three-variable Bernoulli condition with large marginal parameters and time series length of 50, but decreased to 0.602 when time series length was 500. A similar pattern was observed in the mixed distribution conditions. In the Poisson distribution conditions, canonical VAR coverage rates were generally comparable to latent Gaussian rates, although under-coverage emerged when the marginal parameters were small and time series length was large.

\clearpage
\begin{table}[h]
\caption{95\% Confidence Interval Coverage Rate} 
\label{tab:ci}
\centering
{%
\renewcommand{\arraystretch}{1.5}
\scalebox{0.75}{
\begin{tabular}{rrrrrrrrr}
\toprule

&&& \multicolumn{6}{c}{Marginal Distribution}\\
  
\cmidrule(lr){4-9}

&&& \multicolumn{2}{c}{Bernoulli} & \multicolumn{2}{c}{Poisson} & \multicolumn{2}{c}{Mixed} \\

\cmidrule(lr){4-5}\cmidrule(lr){6-7}\cmidrule(lr){8-9} \multicolumn{1}{c}{Variables} &\multicolumn{1}{c}{Time} &\multicolumn{1}{c}{Magnitude} & \multicolumn{1}{c}{Copula} & \multicolumn{1}{c}{VAR} & \multicolumn{1}{c}{Copula} & \multicolumn{1}{c}{VAR} & \multicolumn{1}{c}{Copula} & \multicolumn{1}{c}{VAR}\\

\midrule
3 & 50 & Small & 0.931 & 0.883 & 0.929 & 0.937 & 0.911 & 0.929 \\ 
  3 & 50 & Medium & 0.927 & 0.888 & 0.923 & 0.946 & 0.927 & 0.720 \\ 
  3 & 50 & Large & 0.924 & 0.915 & 0.927 & 0.940 & 0.910 & 0.811 \\ 
  3 & 100 & Small & 0.948 & 0.823 & 0.947 & 0.931 & 0.933 & 0.918 \\ 
  3 & 100 & Medium & 0.941 & 0.812 & 0.941 & 0.943 & 0.937 & 0.666 \\ 
  3 & 100 & Large & 0.942 & 0.868 & 0.937 & 0.940 & 0.936 & 0.753 \\ 
  3 & 200 & Small & 0.948 & 0.708 & 0.954 & 0.911 & 0.944 & 0.898 \\ 
  3 & 200 & Medium & 0.950 & 0.685 & 0.953 & 0.950 & 0.951 & 0.608 \\ 
  3 & 200 & Large & 0.946 & 0.778 & 0.951 & 0.944 & 0.953 & 0.666 \\ 
  3 & 500 & Small & 0.953 & 0.494 & 0.958 & 0.826 & 0.948 & 0.824 \\ 
  3 & 500 & Medium & 0.952 & 0.408 & 0.961 & 0.952 & 0.953 & 0.522 \\ 
  3 & 500 & Large & 0.954 & 0.602 & 0.955 & 0.944 & 0.955 & 0.488 \\ 
  5 & 50 & Small & 0.940 & 0.889 & 0.938 & 0.938 & 0.927 & 0.856 \\ 
  5 & 50 & Medium & 0.927 & 0.901 & 0.943 & 0.944 & 0.933 & 0.767 \\ 
  5 & 50 & Large & 0.940 & 0.897 & 0.933 & 0.944 & 0.930 & 0.624 \\ 
  5 & 100 & Small & 0.955 & 0.822 & 0.953 & 0.930 & 0.944 & 0.780 \\ 
  5 & 100 & Medium & 0.936 & 0.876 & 0.957 & 0.943 & 0.951 & 0.702 \\ 
  5 & 100 & Large & 0.945 & 0.831 & 0.956 & 0.946 & 0.951 & 0.560 \\ 
  5 & 200 & Small & 0.955 & 0.682 & 0.955 & 0.914 & 0.952 & 0.698 \\ 
  5 & 200 & Medium & 0.945 & 0.800 & 0.959 & 0.946 & 0.953 & 0.628 \\ 
  5 & 200 & Large & 0.949 & 0.720 & 0.958 & 0.947 & 0.957 & 0.489 \\ 
  5 & 500 & Small & 0.953 & 0.427 & 0.962 & 0.865 & 0.951 & 0.618 \\ 
  5 & 500 & Medium & 0.952 & 0.634 & 0.965 & 0.943 & 0.958 & 0.495 \\ 
  5 & 500 & Large & 0.950 & 0.520 & 0.965 & 0.949 & 0.959 & 0.415 \\ 
  \bottomrule
  \end{tabular}
}
}
\end{table}

\section{Empirical Example}

We now demonstrate the utility of the derived analytic standard errors using empirical data from Fredrickson et al. (\citeyear{Fredrickson2021}). Data for the current application consisted of two participants who completed daily assessments on a range of constructs across an 11-week period. Three variables were selected from this set to mirror the mixed marginal distributions case detailed in the simulation study. The Bernoulli distributed variable represented engagement in moderate physical activity. More specifically, participants were asked each evening if they had engaged in moderate physical activity (e.g., walking) in the previous 24 hours. The Poisson distributed variable represented the number of alcoholic drinks consumed by participants in the previous 24-hour period. Finally, the Gaussian distributed variable was constructed using positive emotion items (e.g., amused, happy) from the modified differential emotion scale \parencite[mDES;][]{Fredrickson2013}. Participants were asked to consider the degree to which they had experienced each emotion in the prior 24 hours using a 0 (\textit{Not at all}) to 4 (\textit{Extremely}) scale. A composite positive affect (PA) variable was created by averaging the 10 positive emotion items from the mDES for both participants. Consistent with the simulation procedure detailed above, results from the latent Gaussian model were compared with those from a canonical Gaussian VAR.

\begin{figure}
    \centering
    \includegraphics[width=0.8\linewidth]{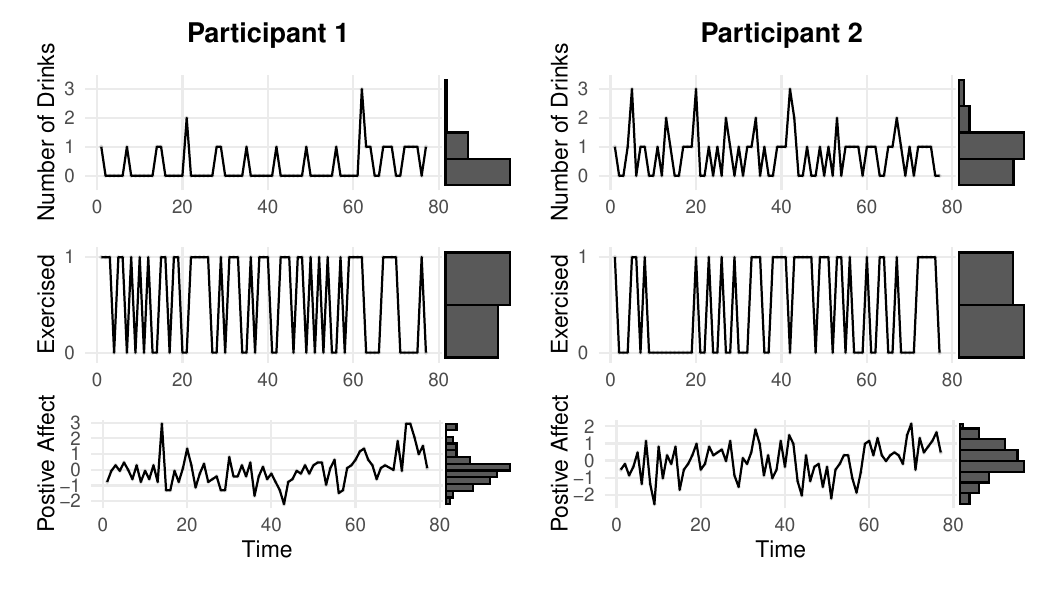}
    \caption{Observed time series and marginal distributions for both participants.}
    \label{fig:raw}
\end{figure}

\begin{figure}
    \centering
    \includegraphics[width=0.8\linewidth]{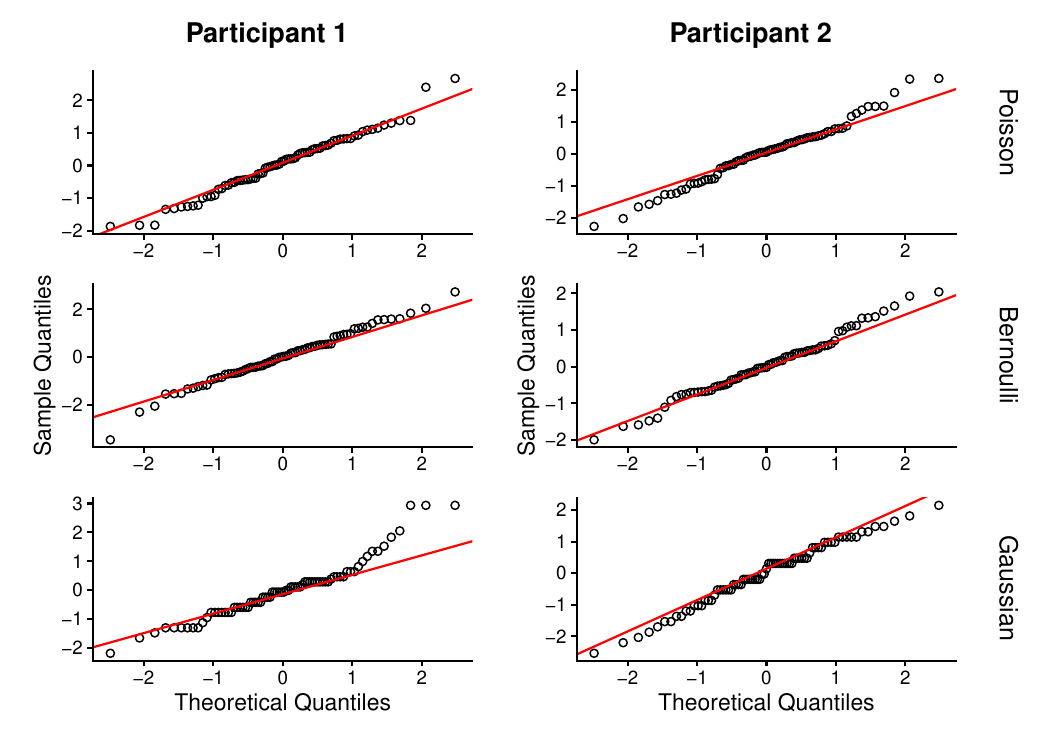}
    \caption{Q--Q plots for observed time series for both participants.}
    \label{fig:qq}
\end{figure}

Prior to estimation of the latent Gaussian dynamics and associated standard errors, we evaluated the plausibility of the distributional and stationarity assumptions described previously. More specifically, the observed time series were visually inspected for any violations of stationarity, such as trends. Deviations from the specified marginal distributions were assessed via quantile--quantile (Q--Q) plots, where the Q--Q plots for the discrete marginal distributions were constructed according to the randomized quantile procedure in Dunn and Smyth (\citeyear{Dunn1996}). As shown in Figure \ref{fig:raw}, pronounced violations of stationarity were not detected. A late-series increase in the PA process was observed for both participants, though this deviation was minor. Similarly, the Q--Q plots shown in Figure \ref{fig:qq} indicate adequate agreement between the observed and specified marginal distributions, with deviations largely confined to the tails. For example, the Gaussian PA series for Participant 1 was characterized by a heavier upper tail than expected, indicating occasional days of high PA endorsement.

Point estimates, estimated standard errors, and 95\% confidence intervals for both participants can be found in Table \ref{tab:empirical}. For Participant 1, both the latent Gaussian and canonical VAR approaches provided evidence for a positive cross-lagged association between PA at day $t$ and number of drinks consumed at day $t+1$. Notably, the latent Gaussian estimate of this dynamic was larger in magnitude $(0.532)$ than the canonical VAR estimate $(0.228)$. Similarly, a larger estimated standard error for the latent Gaussian effect (0.176) resulted in a wider 95\% confidence interval than was observed for the canonical VAR. In addition to the cross-lagged association between PA and number of drinks consumed, the canonical VAR identified nonzero dynamics for both the cross-lagged association between PA at day $t$ and exercise at day $t+1$ and the autoregressive effect of PA at day $t$ and PA at day $t+1$. For Participant 2, both approaches provided evidence for a negative cross-lagged association between number of drinks and subsequent PA, though the estimated dynamic $(-0.297)$ and standard error (0.121) were smaller in the latent Gaussian case. In addition to this effect, the canonical VAR estimated a positive cross-lagged dynamic between PA and drinking behavior, such that an increase in PA was associated with a subsequent increase in the number of drinks consumed.

For both participants, the canonical VAR identified a greater number of effects whose 95\% confidence intervals excluded zero. This pattern is consistent with the simulation study results, in which the canonical VAR exhibited poor nominal 95\% confidence interval coverage under many of the examined conditions. In the mixed-distribution condition, for example, coverage rates ranged from 0.415 to 0.929 for the canonical VAR, whereas coverage rates ranged from 0.910 to 0.959 for the latent Gaussian model. The observed differences in the number of effects deemed nonzero may therefore reflect differences in coverage; that is, the 95\% confidence intervals for the canonical VAR are, on average, too narrow. Consequently, the conventional Wald-based tests and corresponding 95\% confidence intervals for the canonical VAR may exhibit inflated Type I error rates for discrete-valued processes.  

\clearpage
\begin{table}[h]
\caption{Empirical Results}
\label{tab:empirical}
\centering
\setlength{\tabcolsep}{5pt}
\renewcommand{\arraystretch}{1}

\begin{tabular}{ll rrc rrc}
\toprule
& & \multicolumn{3}{c}{Copula} & \multicolumn{3}{c}{VAR} \\
\cmidrule(lr){3-5}\cmidrule(lr){6-8}
 & & \multicolumn{1}{c}{Est} & \multicolumn{1}{c}{SE} & \multicolumn{1}{c}{95\% CI}
  & \multicolumn{1}{c}{Est} & \multicolumn{1}{c}{SE} & \multicolumn{1}{c}{95\% CI} \\
\midrule

\multicolumn{8}{l}{\textbf{Participant 1}}\\
\addlinespace[0.25em]

\multirow{3}{*}{Drinks}
  & Drinks   & -0.084 & 0.125 & \CI{-0.329}{0.161} & -0.010 & 0.111 & \CI{-0.227}{0.207} \\
  & Exercise & -0.297 & 0.296 & \CI{-0.877}{0.283} & -0.191 & 0.124 & \CI{-0.434}{0.053} \\
  & PA       & 0.532 & 0.176 & \textbf{\CI{0.187}{0.878}} & 0.228 & 0.062 & \textbf{\CI{0.107}{0.349}} \\
\addlinespace[0.45em]

\multirow{3}{*}{Exercise}
  & Drinks   & 0.147 & 0.287 & \CI{-0.416}{0.710} & 0.086 & 0.106 & \CI{-0.121}{0.293} \\
  & Exercise & -0.181 & 0.302 & \CI{-0.773}{0.411} & -0.126 & 0.119 & \CI{-0.358}{0.107} \\
  & PA       & -0.353 & 0.180 & \CI{-0.706}{0.001} & -0.127 & 0.059 & \textbf{\CI{-0.242}{-0.011}} \\
\addlinespace[0.45em]

\multirow{3}{*}{PA}
  & Drinks   & -0.094 & 0.166 & \CI{-0.419}{0.231} & -0.112 & 0.207 & \CI{-0.518}{0.294} \\
  & Exercise & -0.025 & 0.191 & \CI{-0.400}{0.350} & -0.014 & 0.233 & \CI{-0.470}{0.442} \\
  & PA       & 0.353 & 0.199 & \CI{-0.037}{0.742} & 0.338 & 0.115 & \textbf{\CI{0.112}{0.565}} \\
\addlinespace[0.8em]

\multicolumn{8}{l}{\textbf{Participant 2}}\\
\addlinespace[0.25em]

\multirow{3}{*}{Drinks}
  & Drinks   & -0.139 & 0.153 & \CI{-0.439}{0.162} & -0.128 & 0.115 & \CI{-0.354}{0.097} \\
  & Exercise & -0.088 & 0.191 & \CI{-0.462}{0.287} & -0.113 & 0.173 & \CI{-0.451}{0.226} \\
  & PA       & 0.301 & 0.176 & \CI{-0.045}{0.647} & 0.231 & 0.087 & \textbf{\CI{0.060}{0.402}} \\
\addlinespace[0.45em]

\multirow{3}{*}{Exercise}
  & Drinks   & 0.125 & 0.147 & \CI{-0.163}{0.414} & 0.064 & 0.077 & \CI{-0.087}{0.215} \\
  & Exercise & 0.198 & 0.176 & \CI{-0.147}{0.542} & 0.125 & 0.116 & \CI{-0.101}{0.352} \\
  & PA       & 0.207 & 0.149 & \CI{-0.085}{0.500} & 0.078 & 0.058 & \CI{-0.037}{0.192} \\
\addlinespace[0.45em]

\multirow{3}{*}{PA}
  & Drinks   & -0.297 & 0.121 & \textbf{\CI{-0.534}{-0.059}} & -0.334 & 0.152 & \textbf{\CI{-0.632}{-0.036}} \\
  & Exercise & 0.042 & 0.144 & \CI{-0.239}{0.324} & 0.049 & 0.228 & \CI{-0.398}{0.496} \\
  & PA       & 0.223 & 0.115 & \CI{-0.002}{0.448} & 0.213 & 0.115 & \CI{-0.012}{0.439} \\

\bottomrule
\end{tabular}
\end{table}

\section{Discussion}

Dynamic processes in the social, behavioral, and health sciences are frequently characterized by discrete-valued variables. Methods for modeling such processes, however, remain underdeveloped, and existing frameworks often lack the flexibility needed to accurately represent the data generating process. The statistical approach detailed herein addresses these challenges by modeling the observed discrete-valued process as a deterministic function of a latent standard Gaussian series. This construction allows for both the accommodation of any discrete-valued marginal distribution and computationally feasible estimation via the second-order properties of the observed series. The current article contributes to the development of this framework by deriving analytic standard errors for Yule--Walker estimates of the latent Gaussian dynamics. The performance of the estimated latent Gaussian dynamics and corresponding standard errors was assessed in a simulation study, and the utility of this approach was demonstrated through an empirical application. Estimates of interest were assessed using the canonical Gaussian VAR as a “naive” alternative.

Results of the simulation study converged to highlight the advantages of the latent Gaussian estimates across a range of conditions. Indeed, though the performance of the latent Gaussian and canonical VAR point estimates was often comparable with respect to absolute bias and RMSE, differences in the performance of the respective estimators emerged when considering 95\% confidence interval coverage rates, especially in the context of processes with mixed marginal distributions. That is, point estimates of the canonical VAR dynamics regularly demonstrated adequate performance, but failed to reach the nominal 0.95 coverage rate. Notably, the observed under-coverage for the canonical VAR estimates became increasingly pronounced at longer time series lengths, indicating that the confidence intervals may be concentrating around a biased estimate. In contrast, coverage rates for the latent Gaussian dynamics exhibited good performance across all conditions. Moreover, estimated standard errors of the latent Gaussian dynamics were generally characterized by low levels of absolute bias, relative bias, and RMSE.

The observed differences in coverage rates were consistent with the results of the empirical application. For both participants, the number of estimated dynamics with 95\% confidence intervals that did not include zero was larger for the canonical VAR approach than for the latent Gaussian model. However, the poor performance of the canonical VAR in the mixed distribution simulation conditions suggests that these nonzero estimates may be a product of model misspecification, such that the estimated intervals are characterized by pronounced under-coverage. Thus, application of the canonical VAR to mixed distribution processes may yield an inflated Type I error rate, resulting in erroneous conclusions regarding the presence of nonzero dynamics. Conversely, the latent Gaussian approach exhibited good performance in these conditions, indicating reliable inference for processes typically encountered in applied settings.

The results of the simulation study and empirical application should be interpreted in the context of several limitations, each of which presents opportunities for future research. First, the latent Gaussian model sometimes produced Yule--Walker estimates which implied nonstationary transition matrices in limited-information conditions. Though the procedure proposed by Higham (\citeyear{Higham2002}) offers one potential remedy to this issue, researchers should exercise caution when employing the latent Gaussian approach in such contexts. For example, processes characterized by short time series lengths and marginal parameters near the boundary of the parameter space may be particularly prone to inadmissible solutions. Future work should explore alternative approaches for approximating the inverse link function in \eqref{eq:inverse}, which may be less likely to produce non-positive definite or ill-conditioned covariance matrices. Alternative post-hoc remedies, such as smoothing procedures proposed in the factor-analytic literature \parencite[e.g.,][]{Lorenzo-Seva2021}, should also be assessed. Next, the estimation procedure described herein cannot currently accommodate missing data. Multivariate extensions of the state-space specification proposed by Jia and colleagues (\citeyear{Jia2023}) could offer one potential remedy. Finally, the construction detailed in \eqref{eq:marginal} ensures that the marginal distributions of the component processes correspond exactly to the specified distribution functions. It is not clear, however, to what extent distributional misspecification impacts the estimation of the second-order properties of the latent Gaussian process. If, for example, an assumed Poisson-distributed process is characterized by overdispersion, the resulting latent series may not be standard Gaussian. Additional research is needed to clarify to what extent the current approach is robust to such misspecifications.

\printbibliography

\appendix

\section{}

The purpose of this appendix is to provide a proof of the main result in \eqref{eq:asymdist} and to derive the exact form of the limiting covariance matrix. The remainder of the appendix is organized as follows. First, we introduce the notation and auxiliary quantities needed to state the limiting covariance matrix compactly. Second, we derive a first-order Taylor expansion of the implied Yule--Walker estimator, which makes explicit the delta-method argument described in the main text. Finally, we use the results of this first-order representation to obtain the limiting covariance matrix. The assumptions required for the proof, which are referred to periodically, are stated in the Supplementary Materials.

\subsection{Notation}

The derivative \(\ell'(u)\) in \eqref{eq:deriv-ell} is defined only for \(u\in(-1,1)\), as $u$ is a latent Gaussian correlation parameter. At the endpoints \(u=\pm1\), the corresponding bivariate Gaussian distribution is degenerate, whereas the derivative formula for the link function contains factors involving \(1-u^2\) and is stated only for \(|u|<1\) (see Proposition~2.1 and Appendix~A.1 in D{\"u}ker et al., \citeyear{Duker2024}). Thus, we separate entries of \(\bm{\Gamma}_Z\) that are fixed at the boundary. Assumption L.1 excludes boundary values for all off-diagonal entries, while the diagonal entries satisfy \(\Gamma_{Z,ii}(0)=1\) since we assume that the latent process is standard Gaussian. We therefore define
\begin{equation} \label{eq:gdot}
g_{\bullet,rs}(v)=
\begin{cases}
1, & \text{if \((r,s)\) corresponds to a variance entry},\\
g_{rs}(v), & \text{otherwise}.
\end{cases}
\end{equation}
This notation will also be used for functions other than $g$ but refers to the same index pairs $r,s$ to be equal to one. We enforce the latent standardization by setting the diagonal (variance) entries of $\widehat g(\acmfXhat)$ equal to $1$ by construction, such that $\widehat{g}( \acmfXhat ) = \widehat{g}_{\bullet}( \acmfXhat )$. This convention fixes the latent variance entries only; all non-variance entries continue to be transformed through the ordinary element-wise inverse link. For instance, we can write
\begin{align} \label{eq:gprime_bullet}
g'_{\bullet,rs}(x)=
\begin{cases}
0, & \text{if \((r,s)\) corresponds to a variance entry},\\
\frac{1}{ \ell'_{rs}(\ell^{-1}_{rs}(x)) }, & \text{otherwise},
\end{cases}
\end{align}
where the derivative is taken with respect to the matrix entry \(x\), holding \(\theta\) fixed.

As established in \eqref{eq:delta1}, the asymptotic behavior of the Yule--Walker estimator is determined by the joint asymptotic distribution of $(\widehat{\theta}, \widehat{\bm{\Gamma}}^{p+1}_{X})$. Thus, our goal is to express $\sqrt{T}(\widehat{\beta} - \beta)$ as a function of the following 
two quantities:
\begin{align} \label{def:Z1Z2}
    Z_{1,T} = \sqrt{T} \vecop ( \widehat{\theta} - \theta ),
    \hspace{0.2cm}
    Z_{2,T} = \sqrt{T} \vecop (\widehat{\bm{\Gamma}}^{p+1}_{X} -  \bm{\Gamma}^{p+1}_{X} ),
\end{align}
which represent estimation error in the marginal parameters and observed autocovariances, respectively. We further define
\begin{align} \label{eq:A1 and A2}
    A_1 = ( I_{\Dim} \otimes \acmfZ^{-1} ) f_1(\theta),
    \hspace{0.2cm}
    A_2 = -
    (\bm{\gamma}_{Z}' \otimes I_{\Dim p} )
    ( (\acmfZ^{-1})' \otimes \acmfZ^{-1} )
    f_2(\theta),
\end{align}
with
\begin{equation} \label{eq:def-f1 and f2}
    f_1(\theta) = \frac{\partial }{\partial \widetilde{\theta} } g_{\bullet}(\gamma_X, \widetilde{\theta}) \bigg\rvert_{\widetilde{\theta} = \theta}
    \hspace{0.2cm}
    \text{ and }
    \hspace{0.2cm}
    f_2(\theta) = \frac{\partial }{\partial \widetilde{\theta} } g_{\bullet}(\acmfX, \widetilde{\theta}) \bigg\rvert_{\widetilde{\theta} = \theta},
\end{equation}
where $f_1(\theta)$ and $f_2(\theta)$ correspond to the derivatives in \eqref{eq:deriv-theta}. A more explicit treatment of these derivatives can be found in the Supplementary Materials. Finally, let
\begin{align*}
    B = I_{\Dim} \otimes \acmfZ^{-1},
    \hspace{0.2cm}
    C = S_3' \otimes S_2' ,
    \hspace{0.2cm}
    D = -
    (\bm{\gamma}_{Z}' \otimes I_{\Dim p} )
    ((\acmfZ^{-1})' \otimes \acmfZ^{-1} ),
\end{align*}
and
\begin{align*}
    E = S_1' \otimes S_1',
    \hspace{0.2cm}
    Q_1 =
    \begin{pmatrix}
        \vecop( g'_{\bullet}( \bm{\gamma}_X ) ) \\
        \vecop( g'_{\bullet}( \acmfX ) )
    \end{pmatrix},
    \hspace{0.2cm}
    Q_2 =
    \begin{pmatrix}
        C \\
        E
    \end{pmatrix},
\end{align*}
where
\begin{equation} \label{eq:AN_eq12}
    S_1' =
    \begin{pmatrix}
        I_{dp} & 0_{dp \times d}
    \end{pmatrix},
    \hspace{0.2cm}
    S_2' =
    \begin{pmatrix}
    J_{dp} & 0_{dp \times d}
    \end{pmatrix},
    \hspace{0.2cm}
    S_3' =
    \begin{pmatrix}
    0_{d \times dp} & I_{d}
    \end{pmatrix}.
\end{equation}
Here, $J_{dp}$ denotes the matrix that results from rotating $I_{dp}$ by 90 degrees, and the derivatives in $Q_1$ are defined element-wise as in \eqref{eq:gprime_bullet}.

\subsection{Asymptotic Linearization of the Yule--Walker Estimator}

We now derive a first-order asymptotic linearization of the Yule--Walker estimator around its population counterpart, thereby expressing the estimator error in terms of \(Z_{1,T}\) and \(Z_{2,T}\). The validity of this linearization relies on the consistency and convergence results for the observed autocovariances established in Section 2 in the Supplementary Materials. We additionally refer to a number of corollaries throughout, which are stated in Section 3 of the Supplementary Materials. We begin by writing $\sqrt{T}(\widehat{\beta} - \beta)$ as a function of the autocovariances of the latent process.
\begin{align}
    \sqrt{T}(\widehat{\beta} - \beta)
    &=
    \sqrt{T}\left(
    ( I_{\Dim} \otimes \widehat{g}( \acmfXhat )^{-1} )
    \vecop( \widehat{g}( \widehat{\bm{\gamma}}_{X} ) )
    -
    ( I_{\Dim} \otimes g( \acmfX )^{-1} )
    \vecop( g( \bm{\gamma}_{X} ) )
    \right)
    \nonumber \\
    &=
    \sqrt{T}\Big(
    ( I_{\Dim} \otimes \widehat{g}( \acmfXhat )^{-1} )
    \vecop( \widehat{g}( \widehat{\bm{\gamma}}_{X} ) )
    -
    ( I_{\Dim} \otimes \widehat{g}( \acmfXhat )^{-1} )
    \vecop( g( \bm{\gamma}_{X} ) )
    \nonumber \\
    &\hspace{1cm}
    +
    ( I_{\Dim} \otimes \widehat{g}( \acmfXhat )^{-1} )
    \vecop( g( \bm{\gamma}_{X} ) )
    -
    ( I_{\Dim} \otimes g( \acmfX )^{-1} )
    \vecop( g( \bm{\gamma}_{X} ) )
    \Big)
    \nonumber \\
    &=
    ( I_{\Dim} \otimes g( \acmfX )^{-1} )
    \sqrt{T}
    \vecop\left(
    \widehat{g}( \widehat{\bm{\gamma}}_{X} )
    -
    g( \bm{\gamma}_{X} )
    \right)
    \nonumber \\
    &\hspace{1cm}
    +
    \left(
    I_{\Dim} \otimes
    \sqrt{T}\left(
    \widehat{g}( \acmfXhat )^{-1}
    -
    g( \acmfX )^{-1}
    \right)
    \right)
    \vecop( g( \bm{\gamma}_{X} ) )
    +
    o_{\Prob}(1)
    \label{eq:AN_eq1} \\
    &=
    ( I_{\Dim} \otimes \acmfZ^{-1} )
    \sqrt{T}
    \vecop\left(
    \widehat{g}( \widehat{\bm{\gamma}}_{X} )
    -
    g( \bm{\gamma}_{X} )
    \right)
    \nonumber \\
    &\hspace{1cm}
    +
    ( g( \bm{\gamma}_{X} )' \otimes I_{\Dim p} )
    \sqrt{T}
    \vecop\left(
    \widehat{g}( \acmfXhat )^{-1}
    -
    g( \acmfX )^{-1}
    \right)
    +
    o_{\Prob}(1),
    \label{eq:AN_eq2}
\end{align}
where \eqref{eq:AN_eq1} follows from Corollary 3.1 since
$\widehat{g}( \acmfXhat ) = \acmfZhat$, and \eqref{eq:AN_eq2} follows from
$\vecop(AB) = (B' \otimes I_m)\vecop(A) = (I_q \otimes A)\vecop(B)$ for matrices $A \in \R^{m\times n},B \in \R^{n\times q}$ \parencite{MagnusNeudecker}.

Focusing on the first summand in \eqref{eq:AN_eq2}, and replacing $g$ and $\widehat{g}$ by $g_{\bullet}$ and $\widehat{g}_{\bullet}$ as in \eqref{eq:gdot}, we have:
\begin{align}
&
    \sqrt{T}\vecop ( \widehat{g}( \widehat{\bm{\gamma}}_{X} ) - g( \bm{\gamma}_{X} ) )
=
    \sqrt{T}\vecop ( 
    \widehat{g}_{\bullet}( \widehat{\bm{\gamma}}_{X} ) 
    - g_{\bullet}( \widehat{\bm{\gamma}}_{X} ))
    + 
    \sqrt{T}\vecop (g_{\bullet}( \widehat{\bm{\gamma}}_{X} ) - g_{\bullet}( \bm{\gamma}_{X} ) )
    \nonumber
\\&=
    \sqrt{T}\vecop ( 
    \widehat{g}_{\bullet}( \bm{\gamma}_{X} ) - g_{\bullet}( \bm{\gamma}_{X} ) )
    + 
    \sqrt{T}\vecop ( (\widehat{g}'_{\bullet}( \bm{\gamma}_{X} ) - g'_{\bullet}( \bm{\gamma}_{X} ) ) \odot (\widehat{\bm{\gamma}}_{X} - \bm{\gamma}_{X} ))
    \nonumber
    \\&\hspace{2cm}+ 
    \sqrt{T}\vecop ( g'_{\bullet}( \bm{\gamma}_{X} ) \odot ( \widehat{\bm{\gamma}}_{X} - \bm{\gamma}_{X}  )) + o_{\Prob}(1)
    \label{eq:AN_eq4}
\\&=
    \sqrt{T}\vecop ( 
    \widehat{g}_{\bullet}( \bm{\gamma}_{X} ) - g_{\bullet}( \bm{\gamma}_{X} ) )
    + 
    \sqrt{T}\vecop (
    g'_{\bullet}( \bm{\gamma}_{X} ) \odot ( \widehat{\bm{\gamma}}_{X} -  \bm{\gamma}_{X} ) ) + o_{\Prob}(1)
    \label{eq:AN_eq5}
\\&=
    \sqrt{T}\vecop ( f_1(\theta) ( \widehat{\theta} - \theta) )
    + 
    \sqrt{T}\vecop (
    g'_{\bullet}( \bm{\gamma}_{X} ) \odot ( \widehat{\bm{\gamma}}_{X} -  \bm{\gamma}_{X} ) ) + o_{\Prob}(1),
    \label{eq:AN_eq6}
\end{align}
where \eqref{eq:AN_eq4} is due to a first-order Taylor expansion of both $\widehat{g}_{\bullet}( \widehat{\bm{\gamma}}_{X} ) - g_{\bullet}( \widehat{\bm{\gamma}}_{X} )$ and $g_{\bullet}$, as well as Lemma 3.1. Then, \eqref{eq:AN_eq5} follows from Corollary 3.2, and \eqref{eq:AN_eq6} results from expanding $g_{\bullet}$ as a function of the marginal parameters.

For the second summand in \eqref{eq:AN_eq2}, we use similar arguments:
\begin{align}
&
    \sqrt{T}\vecop( \widehat{g}( \acmfXhat )^{-1} - g( \acmfX )^{-1} )
=
    \sqrt{T}\vecop( \widehat{g}_{\bullet}( \acmfXhat )^{-1} - g_{\bullet}( \acmfXhat )^{-1}
    + g_{\bullet}( \acmfXhat )^{-1} - g_{\bullet}( \acmfX )^{-1} )
    \nonumber
\\&=
    -\sqrt{T} ( (g_{\bullet}( \acmfXhat )^{-1})' \otimes g_{\bullet}( \acmfXhat )^{-1} )
    \vecop( \widehat{g}_{\bullet}( \acmfXhat ) - g_{\bullet}( \acmfXhat ) )
    \nonumber
    \\&\hspace{1cm}
    -\sqrt{T} ( (g_{\bullet}( \acmfX )^{-1})' \otimes g_{\bullet}( \acmfX )^{-1} )
    \vecop( g_{\bullet}( \acmfXhat ) - g_{\bullet}( \acmfX ) )
     + o_{\Prob}(1)
     \label{eq:AN_eq7}
\\&=
-( (g_{\bullet}( \acmfXhat )^{-1})' \otimes g_{\bullet}( \acmfXhat )^{-1} )
    \sqrt{T} \vecop(  \widehat{g}_{\bullet}( \acmfX ) - g_{\bullet}( \acmfX ) 
    )
    \nonumber
    \\&\hspace{1cm}
    -( (g_{\bullet}( \acmfXhat )^{-1})' \otimes g_{\bullet}( \acmfXhat )^{-1} )
    \vecop( ( \widehat{g}'_{\bullet}( \acmfX ) - g'_{\bullet}( \acmfX )) 
    \odot \sqrt{T} ( \acmfXhat - \acmfX ) )
    \nonumber
    \\&\hspace{1cm}-
    ( (g_{\bullet}( \acmfX )^{-1})' \otimes g_{\bullet}( \acmfX )^{-1} )
    \vecop( g'_{\bullet}( \acmfX ) \odot \sqrt{T} ( \acmfXhat - \acmfX ) )
     + o_{\Prob}(1)
     \label{eq:AN_eq8}
\\&=
-( (g_{\bullet}( \acmfX )^{-1})' \otimes g_{\bullet}( \acmfX )^{-1} )
    \sqrt{T} \vecop(  \widehat{g}_{\bullet}( \acmfX ) - g_{\bullet}( \acmfX ) 
    )
    \nonumber
    \\&\hspace{1cm}
    -
    ( (g_{\bullet}( \acmfX )^{-1})' \otimes g_{\bullet}( \acmfX )^{-1} )
    \vecop( g'_{\bullet}( \acmfX ) \odot \sqrt{T} ( \acmfXhat - \acmfX ) )
     + o_{\Prob}(1)
     \label{eq:AN_eq9}
\\&=
    -
    ( (g_{\bullet}( \acmfX )^{-1})' \otimes g_{\bullet}( \acmfX )^{-1} )
    \sqrt{T}\vecop ( f_2(\theta) ( \widehat{\theta} - \theta) )
    \nonumber
    \\&\hspace{1cm}
    -
    ( (g_{\bullet}( \acmfX )^{-1})' \otimes g_{\bullet}( \acmfX )^{-1} )
    \vecop( g'_{\bullet}( \acmfX ) \odot \sqrt{T} ( \acmfXhat - \acmfX ) )
    + o_{\Prob}(1),
    \label{eq:AN_eq10}
\end{align}
where \eqref{eq:AN_eq7} follows from the vectorized matrix derivative of the inverse function given by
\[
    \bm{d}\vecop\left( (g(A))^{-1} \right)
    =
    -
    \left( 
    ((g(A))^{-1})' \otimes (g(A))^{-1}
     \right)
    \bm{d}\vecop(g(A)).
\]
The $o_{\Prob}(1)$-term is due to Lemmas 3.2 and 3.3. In \eqref{eq:AN_eq8}, a first-order Taylor expansion in $\acmfX$ is applied to $\widehat{g}_{\bullet}( \acmfXhat ) - g_{\bullet}( \acmfXhat )$ in the first summand with the remainder being controlled by Lemma 3.1. Similarly, in the second summand, an expansion is applied to $g_{\bullet}( \acmfXhat )$ at $\acmfX$. Using Corollary 3.2, we get \eqref{eq:AN_eq9}. Finally, we expand $\widehat{g}_{\bullet}$ as a function of the marginal parameters in \eqref{eq:AN_eq10}.

Finally, substituting \eqref{eq:AN_eq6} and \eqref{eq:AN_eq10} into
\eqref{eq:AN_eq2} and noting that 
\begin{align}
    \vecop ( \widehat{\bm{\gamma}}_{X} - \bm{\gamma}_{X} )
    &=
    (S_3' \otimes S_2')
    \vecop\left(
    \widehat{\bm{\Gamma}}_X^{p+1}
    -
    \bm{\Gamma}_X^{p+1}
    \right),
    \label{eq:AN_eq12.2}
    \\
    \vecop ( \acmfXhat - \acmfX )
    &=
    (S_1' \otimes S_1')
    \vecop\left(
    \widehat{\bm{\Gamma}}_X^{p+1}
    -
    \bm{\Gamma}_X^{p+1}
    \right),
    \label{eq:AN_eq12.3}
\end{align}
we obtain
\begin{align}
    \sqrt{T}(\widehat{\beta} - \beta)
    &=
    ( I_{\Dim} \otimes \acmfZ^{-1} )
    (1 \otimes f_1(\theta))
    \sqrt{T}\vecop(\widehat{\theta}-\theta)
    \nonumber \\
    &\hspace{1cm}
    +
    ( I_{\Dim} \otimes \acmfZ^{-1} )
    \left[
    \vecop\left(g'_{\bullet}( \bm{\gamma}_{X} )\right)
    \odot
    (S_3' \otimes S_2')
    \sqrt{T}
    \vecop\left(
    \widehat{\bm{\Gamma}}_X^{p+1}
    -
    \bm{\Gamma}_X^{p+1}
    \right)
    \right]
    \nonumber \\
    &\hspace{1cm}
    -
    ( g_{\bullet}( \bm{\gamma}_{X} )' \otimes I_{\Dim p} )
    \left(
    (g_{\bullet}( \acmfX )^{-1})'
    \otimes
    g_{\bullet}( \acmfX )^{-1}
    \right)
    (1 \otimes f_2(\theta))
    \sqrt{T}\vecop(\widehat{\theta}-\theta)
    \nonumber \\
    &\hspace{1cm}
    -
    ( g_{\bullet}( \bm{\gamma}_{X} )' \otimes I_{\Dim p} )
    \left(
    (g_{\bullet}( \acmfX )^{-1})'
    \otimes
    g_{\bullet}( \acmfX )^{-1}
    \right)
    \nonumber \\
    &\hspace{2cm}
    \times
    \left[
    \vecop(g'_{\bullet}( \acmfX ))
    \odot
    (S_1' \otimes S_1')
    \sqrt{T}
    \vecop\left(
    \widehat{\bm{\Gamma}}_X^{p+1}
    -
    \bm{\Gamma}_X^{p+1}
    \right)
    \right]
    +
    o_{\Prob}(1)
    \label{eq:AN_eq14}
    \\
    &=
    (A_1 + A_2) Z_{1,T}
    +
    [B \ D]
    \left(
    Q_1 \odot (Q_2 Z_{2,T})
    \right)
    +
    o_{\Prob}(1),
    \label{eq:AN_eq16}
\end{align}
where \eqref{eq:AN_eq16} follows from the previously introduced notation and assumptions L.1--L.4, V.1--V.3, and M.1--M.4 in the Supplementary Materials.

\subsection{Limiting Covariance Matrix}

The representation in \eqref{eq:AN_eq16} expresses $\sqrt{T}(\widehat{\beta} - \beta)$ as a linear function of $Z_{1,T}$ and $Z_{2,T}$, up to an asymptotically negligible remainder. Thus, by Slutsky's theorem, the asymptotic behavior and limiting covariance matrix in \eqref{eq:asymdist} follow from the joint asymptotic distribution of $(\widehat{\theta}, \widehat{\bm{\Gamma}}^{p+1}_{X})$ in \eqref{eq:marg-cov dist}. We can therefore derive the limiting covariance matrix $\Sigma_Q$ as
\begin{equation}\label{eq:SigmaQ}
    \begin{aligned} 
    \Sigma_Q
    &=
    \Cov\left( 
    (A_1 + A_2) Z_{1,T}
    +
    [B \ D] \left( Q_1 \odot (Q_2 Z_{2,T}) \right)
    \right)
    \\&=
    (A_1 + A_2) \Sigma_{Z,11} (A_1 + A_2)'
    +
    (A_1 + A_2) \left( J Q_1' \odot \Sigma_{Z,12} Q_{2}'
    \right) [B \ D]'
    \\&\hspace{2cm}+
    [B \ D] \left( Q_1 J' \odot Q_2 \Sigma_{Z,21}
    \right)  (A_1 + A_2)'
    \\&\hspace{3cm}+
    [B \ D] \left( Q_1 Q_1' \odot (Q_2 \Sigma_{Z,22} Q_{2}')
    \right) [B \ D]',
\end{aligned}
\end{equation}
where $J$ is of the same dimension as $Z_{1,T}$ but with all entries equal to $1$.
The explicit block expressions for $\Sigma_Z$, together with the proof of the joint convergence stated in \eqref{eq:marg-cov dist}, are given in the Supplementary Materials.

\end{document}